\documentclass[aps, pra, twocolumn, showpacs,superscriptaddress,10pt]{revtex4-1}
\usepackage{amsfonts,amssymb,amsmath}
\usepackage{graphics,graphicx,epsfig}
\usepackage{amsthm}
\usepackage{color}
\usepackage{epstopdf}

\def\identity{\leavevmode\hbox{\small1\kern-3.8pt\normalsize1}}

\newtheorem{lemma}{Lemma}
\newtheorem{propo}{Proposition}
\newcommand{\be}{\begin{eqnarray}}
\newcommand{\ee}{\end{eqnarray}} 
\newcommand{\bpr}{\begin{propo}}
\newcommand{\epr}{\end{propo}}
\newcommand{\ble}{\begin{lemma}}
\newcommand{\ele}{\end{lemma}}
\newcommand{\bpf}{\begin{proof}}
\newcommand{\epf}{\end{proof}}

\newcommand{\Tr}{\mathrm{Tr}}

\renewcommand{\epsilon}{\varepsilon}
\def\openone{\leavevmode\hbox{\small1 \normalsize \kern-.64em1}}

\usepackage[T1]{fontenc}

\begin{document}

\title{From contextuality of a single photon to realism of an electromagnetic wave}

\author{Marcin Markiewicz} \email{marcinm495@gmail.com}    
\affiliation{Institute of Physics, Jagiellonian University, \L{}ojasiewicza 11,
30-348 Krak\'ow, Poland} 

\author{Dagomir Kaszlikowski}
\email{phykd@nus.edu.sg}
\affiliation{Centre for Quantum Technologies,
National University of Singapore, 3 Science Drive 2, 117543 Singapore,
Singapore}
\affiliation{Department of Physics,
National University of Singapore, 3 Science Drive 2, 117543 Singapore,
Singapore}

\author{Pawe\l{} Kurzy\'nski}   
\email{pawel.kurzynski@amu.edu.pl}
\affiliation{Faculty of Physics, Adam Mickiewicz University, Umultowska 85, 61-614 Pozna\'n, Poland}
\affiliation{Centre for Quantum Technologies,
National University of Singapore, 3 Science Drive 2, 117543 Singapore,
Singapore}

\author{Antoni W\'ojcik}   
\email{antwoj@amu.edu.pl}
\affiliation{Faculty of Physics, Adam Mickiewicz University, Umultowska 85, 61-614 Pozna\'n, Poland}

\date{\today}

%%%%%%%%%%%%%%%%%%%%%%%%%%%%%%%%%%%%%%%%%%%%%%%%%%%%%%%%%%%%%%%%%%%%

\begin{abstract}

Violations of Bell inequalities have been an incontestable indicator of non-classicality since the seminal paper by John Bell. However, recent claims of Bell inequalities violations with classical light have cast some doubts on their significance as hallmarks of non-classicality. Here, we challenge those claims. The crux of the problem is that such classical experiments simulate quantum probabilities with intensities of classical fields. However, fields intensities measurements are radically different from single-photon detections, which are primitives of any genuine Bell experiment. We show that this fundamental difference between field intensities measurements and single photon detections shifts the classical bound of relevant Bell inequalities to its algebraic limit, leaving no place for their violations. 
%This process is an illustration of a quantum-to-classical transition in an experimental correlation-based test, in which the strength of the correlations remains the same for both classical and quantum regimes, while the logical structure of detection events of single outcomes changes from perfect exclusivity in the quantum regime to complete lack of exclusivity in the classical one.

\end{abstract} 
 
%\pacs{...} 

\maketitle 

%%%%%%%%%%%%%%%%%%%%%%%%%%%%%%%%%%%%%%%%%%%%%%%%%%%%%%%%%%%%

The quantum-to-classical transition in optical interferometry can be observed either in a single-particle or multi-particle interference \cite{PanReview}. The multi-particle interference is commonly regarded as more fundamental, including purely quantum phenomena such as the Hong-Ou-Mandel (HOM) effect \cite{HOM} and multi-photon violations of Bell inequalities \cite{LO16}. The quantum-to-classical transition in these phenomena has different origins. In the HOM setting it can be attributed to the strength of particle indistinguishability \cite{Jin17} whereas multi-photon Bell inequality violations are tied to the coherence strength between entangled photons or, in the limit of many particles, to the vanishing ability of revealing single particle properties with multi-photon measurements \cite{Ravi11}. 

%In all the cases, the quantum-to-classical transition manifests in the change of the outcome distribution, like disappearance of the HOM dip or decreasing the strength of correlations in the Bell-type optical tests, leading to lack of a violation in the classical limit. 

A single-photon interference scenario is much simpler to describe. To illustrate it, let us consider the simplest case -- a Young double-slit experiment \cite{SingPhoton}. Here the classical limit is achieved by increasing the average number of photons prepared in a coherent state or a mixture of such states. In this limit, the interference pattern does not change. What changes is the physical meaning of a mathematical formalism used to describe the experiment -- the probability amplitudes of a single photon become amplitudes of a classical electromagnetic wave. Straightforward as it seems, a relatively recent 'discovery' of Bell inequality violations with classical light, dubbed 'classical entanglement', makes the whole classical to quantum transition less obvious.   

Everything started in 1996 with a paper by Patrick Suppes et. al. \cite{Suppes96}. They proposed an interferomteric experiment to violate a Bell inequality with classical light. A year later Robert Spreeuw introduced the concept of classical entanglement between two different degrees of freedom of a single classical light beam \cite{Spreeuw98}. Moreover, he showed that this entanglement leads to violations of the CHSH (Clauser-Horne-Shimony-Holt) inequality \cite{CHSH}. He subsequently generalized this idea to more degrees of freedom, demonstrating a classical version of the GHZ (Greenberger-Horne-Zeilinger) paradox \cite{Spreeuw01}. 

Classical entanglement occurs between two or more properties of an individual system and as such it does not require spatial separation unlike the standard EPR scenario. Because of this, the classical entanglement was largely dismissed by other researchers as a mere curiosity, irrelevant in the context of quantum non-locality \cite{Brunner14}. However a few years later a variety of papers appeared, discussing a similar concept of \emph{intrasystem entanglement} in different physical implementations. Eberly, Qian \emph{et. al.} and further Aiello \emph{et.al} in a series of papers \cite{Eberly2011, Eberly2013, Aiello15NJP} developed a theory of "bipartite" entanglement between polarization and position degrees of freedom in stochastic light beams \cite{Wolf07}. Later, Eberly {\it et. al.} performed an experiment achieving a strong violation of the CHSH inequality with entangled states of stochastic light fields \cite{Eberly2015}. A similar violation of the CHSH inequality with \emph{classical entanglement} between fields in two optical resonators was proposed by Snoke \cite{Snoke14}. Finally Frustaglia \emph{et. al.} \cite{CabelloSimulations} derived a procedure, following earlier ideas of Cerf \emph{et. al.} \cite{Cerf98} and Spreeuw \cite{Spreeuw01}, which allows to reconstruct probability distributions coming from any quantum correlations tests using classical optical circuits. As an illustration of their method, the authors of \cite{CabelloSimulations} performed an experiment with microwave circuits showing violations of the CHSH \cite{CHSH} and Mermin \cite{Mermin90} type inequalities.

Experimental demonstrations of Bell inequalities violations with classical light have profound physical implications. Snoke \cite{Snoke14} and Qian \emph{et.al.} \cite{Eberly2015} hypothesised that a Bell inequality violation does not testify a presence of quantum entanglement in a given physical system -- it may as well be a classical entanglement. Another hypothesis by Frustaglia \emph{et. al.} \cite{CabelloSimulations} is that the bounds on the strength of quantum correlations (so called \emph{Tsirelson bounds}) are not restricted to quantum physics, but arise naturally in classical systems which simulate quantum correlations. If these claims were true, we would have to reconsider the role of Bell inequalities in probing quantum to classical transition.  

In this paper we challenge these claims by showing that one does not observe any violation in classical regime if the Bell inequalities are properly derived. More precisely, Bell inequalities test if probability distributions of measurement results are contextual \cite{KS} -- the feature commonly accepted as an indicator of non-classicality. From the mathematical perspective, Bell inequalities are based on properties of exclusivity relations between jointly measurable events \cite{CSW}. A proper structure of such exclusivity implies existence of a test that can distinguish between contextual (non-classical) and non-contextual (classical) probability distributions. 
%We show that a necessery condition for the contextuality of a probability distribution modelling some experiment is that the events representing different possible outcomes of joint measurements for  fixed settings must be exclusive. 

%We show that in so far proposed experiments the exclusivity structure of measurable events leads to Bell type inequalities that can never be violated because the corresponding classical bound is equal to the algebraic bound. This result holds for any system composed of many indistinguishable subsystems, such as electromagnetic waves. Therefore, we prove that such systems are noncontextual with respect to the proposed tests.
%whose this condition is never fulfilled by the Bell type tests based on classical entanglement present in classical optical waves in so far proposed experiments. 

We show a non-contextual physical model based on quantum-to-classical transition from single photons to classical waves. Our model proves that classical waves are not contextual and thus they can be still called classical. Moreover, we demonstrate that the proper classical bounds, i.e., the bounds respecting the correct exclusivity structure of detection events for Bell tests with classical light are equal to the algebraic bounds on the correlations' strength. Therefore, there is no place for any \emph{classical contextuality} in such systems.

% We also explain why in the experiments with classical light the maximal strength of correlations is exactly as the quantum Tsirelson bound. {\bf We show that this remarkable coincidence is a consequence of dynamical constraints on the unitary evolution of classical waves and it is not related to the long-standing discussion of what physical principles bound the strength of quantum correlations - DISCUSSION}. 

%%%%%%%%%%%%%%%%%%%%%%%%%%%%%%%%%%%%%%%%%%%%%%%%%%%%%%%%%%%%

\section{Results}

%%%%%%%%%%%%%%%%%%%%%%%%%%%%%%%%%%%%%%%%%%%%%%%%%%%%%%%%%%%%

\subsection{Photon distribution in the classical limit}

Consider a single photon in a polarisation state $\sqrt{p_H}|H\rangle + \sqrt{p_V}|V\rangle$, where $H$ and $V$ denote horizontal and vertical polarisations, respectively, and $p_H+p_V=1$. When incident on a polarising beam splitter (PBS), the photon can either go through and become $H$ polarised, or be reflected and become $V$ polarised. These are two possibilities occuring randomly with probabilities $p_H$ and $p_V$, if one decides to detect the photon after the PBS. Denote these two possible outcomes as $\{0,1\}$ and $\{1,0\}$. This scenario is a physical implementation of a binary $\pm 1$ random variable $X$, where the outcome $\{1,0\}$ is associated with the value $+1$ and $\{0,1\}$ with $-1$. 

Next, let us consider two indistinguishable photons in the above state entering the same PBS port. They are uncorrelated and therefore they scatter on the PBS independently \cite{Loudon}. The photons cannot be distinguished and therefore there are only three exclusive outcomes: $\{2,0\}$, $\{1,1\}$ and $\{0,2\}$. These outcomes cannot be interpreted as products of two single-photon outcomes becasue of indistinguishability, i.e., events $\{1,0\}\times \{1,0\}$, $\{1,0\}\times \{0,1\}$, $\{0,1\}\times \{1,0\}$ and $\{0,1\}\times \{0,1\}$ are meaningless. Moreover, unlike in the single-photon case, for two photons, statements "photon is detected on the left" and "photon is detected on the right" are not exclusive because there is a chance that photons can be detected on both sides. Interestingly, the average number of photons in each output is proportional to single-photon scattering probabilities, i.e., $\bar{n}_H=2p_H$ and $\bar{n}_V=2p_V$. 

In general, for $N$ photons scattering on the PBS one can observe $N+1$ exclusive outcomes: $\{N,0\}$, $\{N-1,1\}$, etc. This is drastically fewer than $2^N$ outcomes observable for distinguishable particles. For $N$ photons the single-photon random variable $X$ is ill-defined because of indistinguishability. However, it is possible to define a random variable whose outcomes are given by ${\cal X}=(n_H-n_V)/N$, i.e., the difference between photon numbers in the output ports divided by the total number of photons. Note, that $-1\leq {\cal X}\leq 1$ and ${\cal X}=X$ for $N=1$. Additionally, since each photon is transformed independently and according to the same rule, the average number of photons in each polarisation mode is given by  $\bar{n}_H=Np_H$ and $\bar{n}_V=Np_V$. Because of this $\langle {\cal X} \rangle=p_H - p_V$ does not depend on $N$ and the most probable outcome state is $\{\bar{n}_H,\bar{n}_V\}$.

Finally, let us discuss the classical limit. In this case the total number of photons is undetermined but their average number is large ($\langle N \rangle \gg 1$). In quantum theory such situations are usually represented by a high amplitude coherent state \cite{QOpticsbook}. Once we go to the classical limit, it is quite natural to treat the beam of light as a continuous object that can be split into portions in an arbitrary way. The PBS transforms a single beam with intensity $I$ into two beams, the $H$ polarised beam with intensity $I_H$ and the $V$ polarised one with $I_V$. This is predicted by both, classical and quantum theories. In the classical limit the average value of random variable ${\cal X}$ becomes $(I_H - I_V)/I$. However, since the intensities of two beams are given by $I_H=Ip_H$ and $I_V=Ip_V$, we get $\langle {\cal X} \rangle = p_H-p_V$, as expected. Note, that the fluctuations of ${\cal X}$ scale as $1/\sqrt{N}$, therefore in the classical limit ${\cal X}$ can be treated as a deterministic variable.  

%%%%%%%%%%%%

\begin{figure}[t]
\includegraphics[width=0.45 \textwidth,trim=4 4 4 4,clip]{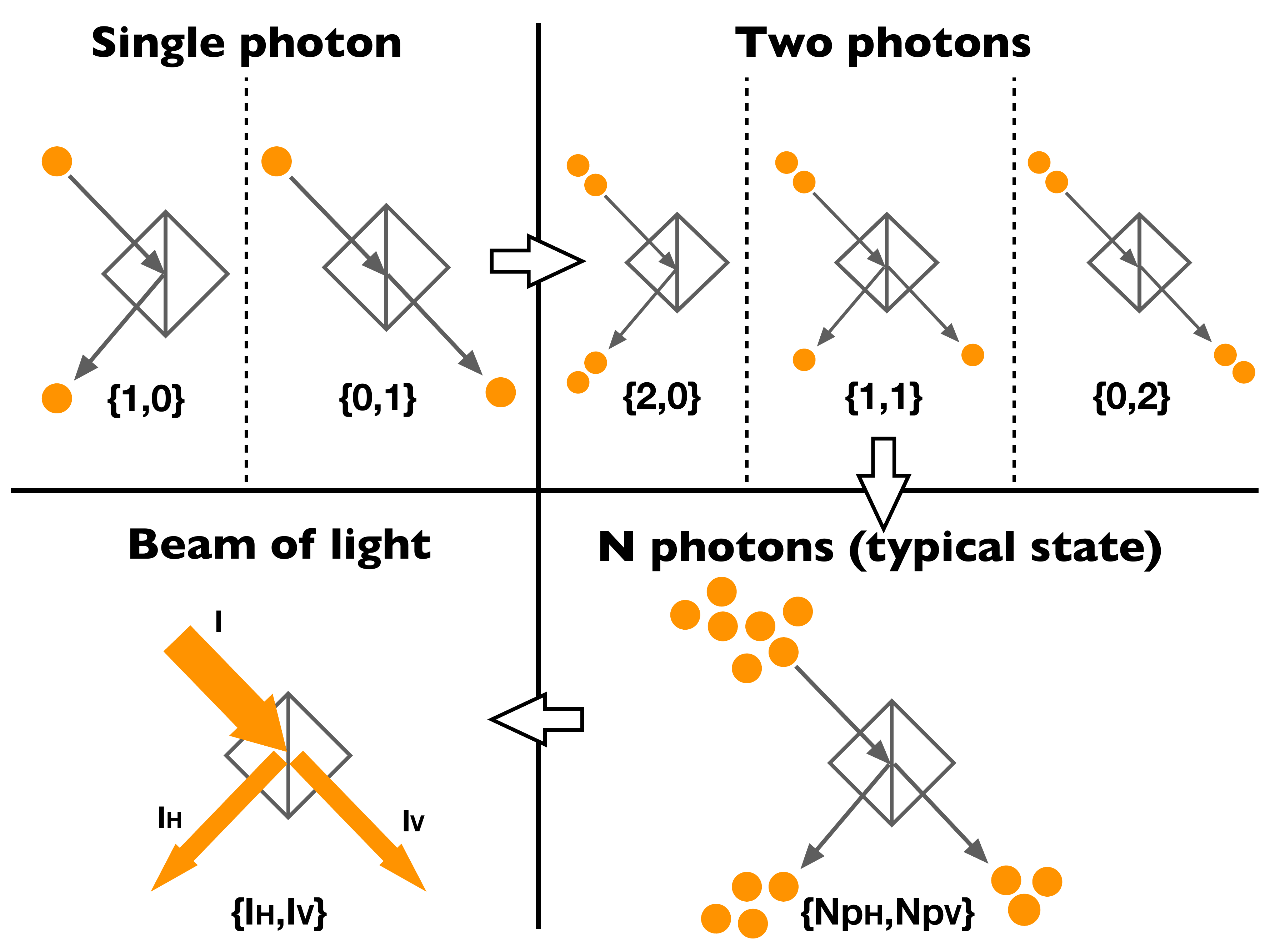}
\caption{ Schematic representation of the classical limit in an experiment with uncorrelated photons. Single photon on a polarising beam splitter (PBS) can either go through or reflect. There are two exclusive outcomes: the photon is either registered on the left with the polarisation $V$ or on the right with the polarisation $H$. The corresponding probabilities are $p_V$ and $p_H$, respectively ($p_H+p_V=1$). Two photons on the PBS can produce three exclusive outcomes: both on the left with probability $p_V^2$, one on the left and one on the right with probability $2 p_H p_V$, and both on the right with probability $p_H^2$. $N$ photons on the PBS can produce $N+1$ exclusive outcomes, however the most probable events are those  with approximately $N p_V$ photons on the left and $N p_H$ photons on the right. A classical beam of light of intensity $I$ is split on the PBS into two beams. In principle there is a continuum of outcomes, however one always observes the one with the corresponding intensities $I_H$ and $I_V$, where $I_H=I p_H$ and $I_V=I p_V$.}
\label{fig1}  
\end{figure}

%%%%%%%%%%%%

The above scenarios are schematically represented in Fig. \ref{fig1}. Although we considered only a single PBS, the similarity between classical intensities and probabilities generated by photonic distributions would also hold if one used an arbitrary number of linear optical devices (PBS, standard beam splitters (BS), phase shifters, etc). In this case the whole setup is equivalent to a multiport corresponding to a more complex random variable or a sequence of random variables.   

To summarise, we see that the same average value $\langle {\cal X}\rangle$ is predicted by both, quantum and classical theories, since this value does not depend on $N$. Nevertheless, the underlying exclusivity structure of the outcomes of ${\cal X}$ strongly depends on $N$. This fact causes some bizarre interpretation difficulties in experimental Bell-type scenarios with classical light. We will discuss this problem in more details in the following sections.

%%%%%%%%%%%%%%%%%%%%%%%%%%%%%%%%%%%%%%%%%%%%%%%%%%%%%%%%%%%%

\subsection{Correlations in the classical limit}

Next, we show that unlike photonic distribution, the correlations between spatially separated photons strongly depend on $N$ and on the exclusivity structure of detection events. The problem of quantum correlations in the classical limit was discussed in details before (see for example \cite{Ravi11}) so we provide here only a simple example. 

Consider a pair of photons in an entangled polarisation state $\sqrt{p_H}|HH\rangle + \sqrt{p_V}|VV\rangle$. These two photons are shared between two spatially separated parties, Alice and Bob, who measure their photons polarisations with respective PBSs. As before, we represent the two polarisation possibilities by $\{1,0\}$ and $\{0,1\}$. Moreover, we can also use the $\pm 1$ random variables $X_A$ and $X_B$, defined in the same way as $X$ above, to represent the measurement of Alice and Bob. Alice and Bob register either $\{1,0\}\times\{1,0\}$ with probability $p_V$, or $\{0,1\}\times\{0,1\}$ with probability $p_H$. The average values and the corresponding correlations are $\langle X_A  \rangle=\langle X_B\rangle = p_H - p_V$ and $\langle X_A X_B \rangle = 1$. 

Next, let us consider $N$ such photonic pairs shared between Alice and Bob who measure polarisation on all pairs at the same time. As before, the local measurements are represented by the random variables ${\cal X}_A=(n_{AH}-n_{AV})/N$ and ${\cal X}_B=(n_{BH}-n_{BV})/N$. Again, there are correlations between the photonic pairs giving $n_{AH}=n_{BH}=n_{H}$ and $n_{AV}=n_{BV}=n_{V}$, hence Alice registers the same photon distribution as Bob and ${\cal X}_A={\cal X}_B$. However, for $N$ entangled pairs the correlations $\langle {\cal X}_A {\cal X}_B\rangle$ are much weaker. Note, that the outcome $\{n,N-n\}\times\{n,N-n\}$ happens with probability $\frac{N!}{n!(N-n)!}p_V^n p_H^{N-n}$, thus
\begin{eqnarray}
\langle {\cal X}_A {\cal X}_B\rangle &=& \sum_{n=0}^N \left(\frac{2n-N}{N}\right)^2 \frac{N!}{n!(N-n)!}p_V^n p_H^{N-n}  \nonumber \\
&=& \frac{N - 4p_Vp_H (N-1)}{N}.
\end{eqnarray}
For example, for $p_H=p_V=1/2$ one gets $\langle {\cal X}_A {\cal X}_B\rangle=1/N$. Interestingly, $\langle {\cal X}_A \rangle = \langle {\cal X}_B \rangle= p_H - p_V$ and in the limit of the large number of photons
\begin{equation}
\lim_{N\rightarrow \infty}\langle {\cal X}_A {\cal X}_B\rangle = 1 -4p_Vp_H =  \langle {\cal X}_A \rangle\langle {\cal X}_B \rangle.
\end{equation}

To conclude, the values $\langle {\cal X}_A \rangle$ and $\langle {\cal X}_B \rangle$ do not depend on $N$. However, the correlation between ${\cal X}_A$ and $ {\cal X}_B$, $\langle {\cal X}_A {\cal X}_B\rangle$, does. As a consequence, in the classical limit of large $\langle N \rangle$ the two random variables get practically uncorrelated. Therefore, the classical limit of Bell-type scenarios based on correlations between many particles can always be explained by a classical theory (for more details see \cite{Ravi11} and the methods). 
To reinforce our statement let us say that correlations between individual photons cannot be used to mimic any non-classical correlations in the limit of classical beams. The idea of a classical simulation of quantum correlations using classical beams uses different approach, and in the next section we focus on correlations between random variables defined for the same particle.

%%%%%%%%%%%%%%%%%%%%%%%%%%%%%%%%%%%%%%%%%%%%%%%%%%%%%%%%%%%%

\subsection{Bell inequalities in the clasical limit}

Let us consider the CHSH scenario \cite{CHSH}, which is the simplest Bell test involving four $\pm 1$ binary random variables $A_0$, $A_1$, $B_0$ and $B_1$. In a classical theory these four random variables are jointly distributed and the following inequality must be satisfied
\begin{equation}\label{CHSH}
-2 \leq \langle A_1 B_1\rangle + \langle A_0 B_1\rangle + \langle A_1 B_0\rangle - \langle A_0 B_0\rangle \leq 2,
\end{equation}
where 
\begin{equation}
\langle A_i B_j\rangle = \sum_{a_i,b_j=\pm 1} a_i b_j p(A_i=a_i,B_j=b_j).
\end{equation}

In quantum theory it is possible to find a set of binary observables represented by Hermitian matrices, such that $[A_i,B_j] \equiv A_i B_j - B_j A_i =0$ (for $i,j=0,1$), but $[A_0,A_1]\neq 0$ and  $[B_0,B_1]\neq 0$. This means that $A_i$ and $B_j$ can be jointly measured, but it is not possible to jointly measure $A_0$ and $A_1$ or $B_0$ and $B_1$. Interestingly, for quantum correlations $\langle A_i B_j \rangle$ the inequality (\ref{CHSH}) can be violated up to $\pm 2\sqrt{2}$ for an optimal choice of the state and observables. The violation implies that the measured correlations cannot be described by classical theories.

\begin{figure}[t]
\includegraphics[width=0.45 \textwidth,trim=4 4 4 4,clip]{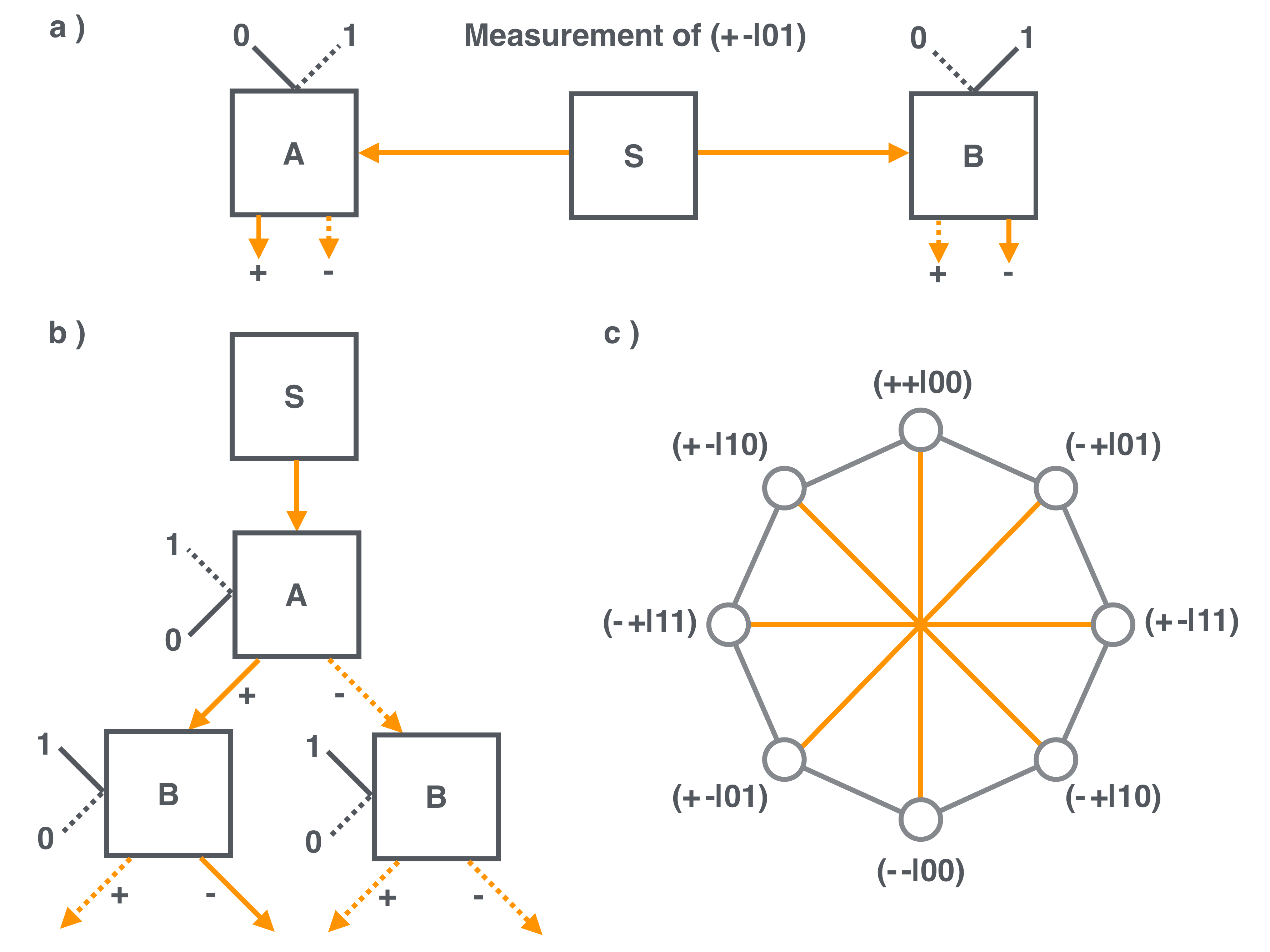}
\caption{Clauser-Horne-Shimony-Holt (CHSH) Bell-type scenario. a) Nonlocal setting in which a source S emits two correlated particles flying to Alice and Bob. Each of them performs one of the two measurements (denoted by 0 or 1).  Each measurement produces a binary outcome ($\pm 1$). Here, we present an instance in which Alice chooses to measure 0 and Bob chooses to measure 1. Alice's outcome is $+$ and Bob's is $-$, hence they jointly register an event $(+-|01)$. b) The same instance, but in a local scenario. Classical entanglement can only be tested in such scenarios. The measurement of $A$ is performed before $B$. It is generally assumed that both properties are compatible (they commute in QM sense), therefore the order of measurement is irrelevant. c) The exclusivity graph for the CHSH scenario. The vertices correspond to measurement events and the edges represent the exclusivity relations. Orange edges correspond to exclusivity of measurement outcomes for the same settings, e.g. $(++|ij)$ and $(--|ij)$. Grey edges correspond to exclusivity of measurement outcomes in which the second measurement has different settings, e.g. $(++|i0)$ and $(--|i1)$. This exclusivity can be tested in the setting represented in b), by choosing $B_0$ for the second left measuring device and $B_1$ for the second right measuring device -- detailed discussion in the text.}  
\label{fig3}
\end{figure}

The simplest quantum system where such a scenario is possible has four levels. In the original Bell-type scenario we have two spatially separated systems, e.g. two polarisation entangled photons discussed in the previous section, see also Fig. \ref{fig3} a). In this case $A_0$ and $A_1$ correspond to the polarisation properties of the first photon, whereas  $B_0$ and $B_1$ correspond to the polarisation properties of the second photon. However, using the arguments from the previous section,  a large number of indistinguishable entangled pairs would produce $\langle A_i B_j \rangle \approx \langle A_i\rangle \langle B_j \rangle$ in the classical limit. Thus, the CHSH inequality (\ref{CHSH}) would not be violated.

%%%%%%%%%%%%

%%%%%%%%%%%%

Let us now discuss another implementation of the CHSH scenario. This time the four-level system is made of a single photon which can occupy four modes, e.g., two polarisation modes ($H$ and $V$) and two spatial modes ($a$ and $b$). As a result the photon can be in one of four possible states $a_H$, $a_V$, $b_H$ and $b_V$, or in an arbitrary superposition of them. The properties $A_0$ and $A_1$ can be associated with spatial modes, whereas $B_0$ and $B_1$ can be associated with polarisation. For example, $A_0$ can assign $+1$ to mode $a$ and $-1$ to $b$. On the other hand, $A_1$ can assign $\pm1$ to orthogonal superpositions of modes, like $|a\rangle \pm |b\rangle$. Similarly, $B_0$ can assign $+1$ to polarisation $H$ and $-1$ to $V$, whereas $B_1$ can assign $+1$ to the right-handed circular polarisation and $-1$ to the left-handed one. 

The Hilbert space of the system is a tensor product of two Hilbert spaces: the one corresponding to spatial modes and another one to polarisation. However, this time the system cannot be divided into parts that can be separated from each other. Still, it is possible to speak of entanglement between these two degrees of freedom, but this entanglement has nothing to do with nonlocality. Nevertheless, violation of the CHSH inequality with $A_i$ and $B_j$ confirms the presence of entanglement between spatial modes and polarisation. This entanglement gives non-classical correlations that can be attributed to contextuality rather than to nonlocality.

The properties $A_i$ and $B_j$ can be measured sequentially, as in \cite{CabelloSimulations}, and the measurement of one property does not disturb the measurement of the other. More precisely, such a measurement can be implemented in a setup in which the system goes through the measuring device corresponding to $A_i$ and then through one of the two measuring devices corresponding to $B_j$. The schematic representation of this setup is shown in Fig. \ref{fig3} b). Because $A_i$ and $B_j$ commute, the results of the measurements do not depend on their order, i.e., $B_j$ can be measured before $A_i$. The measurements lead to four possible outcomes that we denote by $(++|ij)$, $(+-|ij)$, $(-+|ij)$ and $(--|ij)$. The result $(+-|ij)$ corresponds to $A_i=+1$ and $B_j=-1$. 

A single run of the experiment makes one of the four detectors, placed after the outputs, click. The probabilities of these clicks are $p(++|ij)$, $p(+-|ij)$, $p(-+|ij)$ and $p(--|ij)$. They can be estimated after many experimental runs and used to evaluate correlations $\langle A_i B_j\rangle = p(++|ij)-p(+-|ij)-p(-+|ij)+p(--|ij)$. One can observe violation of the CHSH inequality if in each experimental run the photon is prepared in the same special state and the measurements $A_i$ and $B_j$ are properly chosen. Although the setup is interpreted as a measurement of two random variables, it can also be viewed as a measurement of a single degenerate random variable $X_{ij}$ whose outcomes are products of the outcomes of $A_i$ and $B_j$. Therefore, $\langle A_i B_j \rangle = \langle X_{ij} \rangle$.

What would happen if in a single experimental run one used many identical photons or a classical beam of light? From our initial discussion we know that the intensities at the outputs would be proportional to $Np(++|ij)$, $Np(+-|ij)$, $Np(-+|ij)$ and $Np(--|ij)$, where $N$ is the number of photons. In the classical limit one would deal with a beam of light whose intensities would be $I(++|ij)$, $I(+-|ij)$, $I(-+|ij)$ and $I(--|ij)$. Moreover, $I(++|ij)/I=p(++|ij)$, etc., where $I$ is the input intensity. 

In addition, one could consider a random variable 
\begin{equation}
{\cal X}_{ij} = \frac{n(++|ij) - n(+-|ij) - n(-+|ij) + n(--|ij)}{N},
\end{equation}
where $n(++|ij)$ is the number of photons in the output $(++|ij)$, etc. For a single photon $\langle A_i B_j \rangle =\langle X_{ij} \rangle = \langle {\cal X}_{ij} \rangle$.  For $N>1$ it is impossible to assign definite values to $A_i$, $B_j$ and to assign $X_{ij}$ to individual photons. However, $\langle {\cal X}_{ij} \rangle$ can be evaluated and in the classical limit one gets
\begin{equation}\label{classical average}
\langle {\cal X}_{ij} \rangle = \frac{I(++|ij) - I(+-|ij) - I(-+|ij) + I(--|ij)}{I}.
\end{equation}
Thus, it is possible to prepare a classical state of light such that
\begin{equation}
\langle {\cal X}_{11}\rangle + \langle {\cal X}_{01}\rangle + \langle {\cal X}_{10}\rangle - \langle {\cal X}_{00}\rangle = \pm 2\sqrt{2}.
\end{equation}

The above may lead to a discussion whether the classical light has some nonclassical properties \cite{Suppes96,Spreeuw98,Spreeuw01,Eberly2011, Eberly2013, Aiello15NJP,Eberly2015,Snoke14,CabelloSimulations}. In the following sections we show that for more than one photon the classical bound is different than $\pm 2$. One needs to remember that although $\langle {\cal X}_{ij} \rangle$ does not depend on $N$, the random variable $ {\cal X}_{ij}$ and the corresponding exclusivity structure of events strongly depends on $N$, therefore in order to understand what is really going on it is better to examine the CHSH scenario from the point of view of events, not averages.

%%%%%%%%%%%%%%%%%%%%%%%%%%%%%%%%%%%%%%%%%%%%%%%%%%%%%%%%%%%%

\subsection{Exclusivity and classical bounds}

The CHSH inequality can be rewritten with probabilities of detection events. Since $\langle A_i B_j\rangle  = 1 - 2p(+-|ij) - 2p(-+|ij) = 2p(++|ij) + 2p(--|ij) - 1$, the inequality (\ref{CHSH}) becomes 
\begin{eqnarray}
p(+-|11)+p(-+|11)+p(+-|01)&+& \nonumber \\ p(-+|01) + p(+-|10)+p(-+|10) &+& \nonumber \\ p(++|00)+p(--|00) &\leq& 3.  \label{CHSH2}
\end{eqnarray}
This inequality can be derived in a completely different way. The upper bound equal to three comes from the exclusivity structure of events. Firstly, the events $(++|ij)$, $(--|ij)$, $(+-|ij)$ and $(-+|ij)$ are pairwise exclusive. This is because they correspond to different outcomes of the same measurements. For example, $(++|00)$ cannot happen together with $(--|00)$. In addition, two events are exclusive if they share the same measurement settings and the corresponding outcomes are different. This means that $(+\#|ij)$ is exclusive to  $(-\#|ik)$ and $(\#+|ij)$ is exclusive to $(\# -|kj)$; Here $\#$ denotes an arbitrary outcome. For example, $(+-|10)$ is exclusive to $(-+|11)$ and $(--|00)$ is exclusive to $(-+|10)$. Such example can be realised in quantum theory by events corresponding to projections onto states $|0\rangle\otimes|0\rangle$ and $|1\rangle\otimes(\alpha |0\rangle + \beta|1\rangle)$. Although $|0\rangle$ and $\alpha|0\rangle + \beta|1\rangle$ are in general nonorthogonal states, the exclusivity is provided by the orthogonality of $|0\rangle$ and $|1\rangle$ in the first Hilbert space. Verification of this type of exclusivity can be implemented in the sequential scenario represented in Fig. \ref{fig3} b) in which the second left measuring device is set to $B_0$ and the second right to $B_1$.

The exclusivity structure of the eight events can be represented with the {\it exclusivity graph} \cite{CSW} whose vertices correspond to events and edges to exclusivity between two events, see Fig. \ref{fig3} c). The upper bound of (\ref{CHSH2}) is derived under assumption that the eight events are jointly distributed \cite{Fine82}. The joined probability distribution (JPD) is constructed over all possible assignments of $1/0$ (truth/false) values to these events. In principle there are $2^8$ possible assignments, however the value $1$ cannot be simultaneously assigned to two exclusive events. This significantly reduces the number of possible assignments. The maximum value of the sum of the eight probabilities is given by the maximal number of events that can be assigned the value $1$. The problem of finding the maximal number of events that can be assigned $1$ is equivalent to the graph theoretical problem known as {\it maximum independent set} \cite{CSW}. An independent set of a graph is a set of disconnected vertices. We are looking for a set with the largest possible number of vertices. In general, it is an NP-hard problem but it is solvable for our graph. Note, that the set of events that are assigned $1$ must correspond to the independent set of the exclusivity graph, since two events from such set cannot be exclusive. It is easy to find that the maximum independent set of the graph from Fig. \ref{fig3} c) contains three vertices. Therefore, the sum of the eight probabilities cannot be larger than three if these probabilities originate from some JPD. Not surprisingly, quantum theory can go as high as $2+\sqrt{2}$ and it cannot be modelled with any JPD. 

%%%%%%%%%%%%%%%%%%%%%%%%%%%%%%%%%%%%%%%%%%%%%%%%%%%%%%%%%%%%

\subsection{(Non-)contextuality of many indistinguishable particles and a proper classical bound}

The $1/0$ assignment corresponds to a deterministic non-contextual (NC) model. The photon is assigned at most one event from each set of pairwise exclusive events. Such a set makes a measurement context -- a set of events that can be jointly measured. If a context is complete, i.e., it consists of all possible measurement outcomes, the photon is assigned exactly one event. However, in the scenario considered here all contexts are not complete and contain exactly two events.  

To properly discuss the problem of non-classicality of correlations in Bell-type scenarios for classical light, we need to redefine the introduced exclusivity graph model so that a transition from a single photon to a macroscopic electromagnetic wave is transparent. Instead of assigning events to a photon one should tie a photon to an event. This is a subtle difference but it leads to fundamental consequences once we deal with more than one photon. More precisely, a photon is assigned to at most one single event in each measurement context, where $1$ corresponds to a photon and $0$ to no photon event. In this new picture the events can be considered as modes and $1/0$ as occupation numbers. The NC model assigns a well defined occupation number to each mode. The exclusivity leads to conservation of the particle number -- since there is a single photon in the system there could be at most a single photon in each context. If two exclusive events were assigned one, then there would exist a context containing two photons, which would contradict conservation of the particle number. The above interpretation was proposed for the first time in \cite{Kurzynski2017}. This approach is discussed in details in the Methods section. It should be emphasised that the introduced model is very general and describes the single-photon-to-classical-wave transition in Bell-type scenarios irrespective of the direct physical implementation, which may be introduced in many different scenarios \cite{Suppes96, Cerf98,Spreeuw01, CabelloSimulations, Snoke14, Eberly2015, Aiello15NJP}.

One can now rewrite the inequality (\ref{CHSH2}) as
\begin{eqnarray}
\langle n(+-|11)\rangle +\langle n(-+|11)\rangle +\langle n(+-|01)\rangle &+& \nonumber \\ \langle n(-+|01)\rangle + \langle n(+-|10)\rangle +\langle n(-+|10)\rangle &+& \nonumber \\ \langle n(++|00)\rangle +\langle n(--|00) \rangle &\leq& {\cal C},  \label{CHSH3}
\end{eqnarray}
where $n(++|ij)$, etc., are occupation numbers of the corresponding events and ${\cal C}$ is the NC bound on the sum of these numbers, which in the case of a single photon equals to three. A single photon violates this bound.

%%%%%%%%%%%%%%%%%%%%%%%%%%%%%%%%

\begin{figure}[t]
\includegraphics[width=0.5 \textwidth,trim=4 4 4 4,clip]{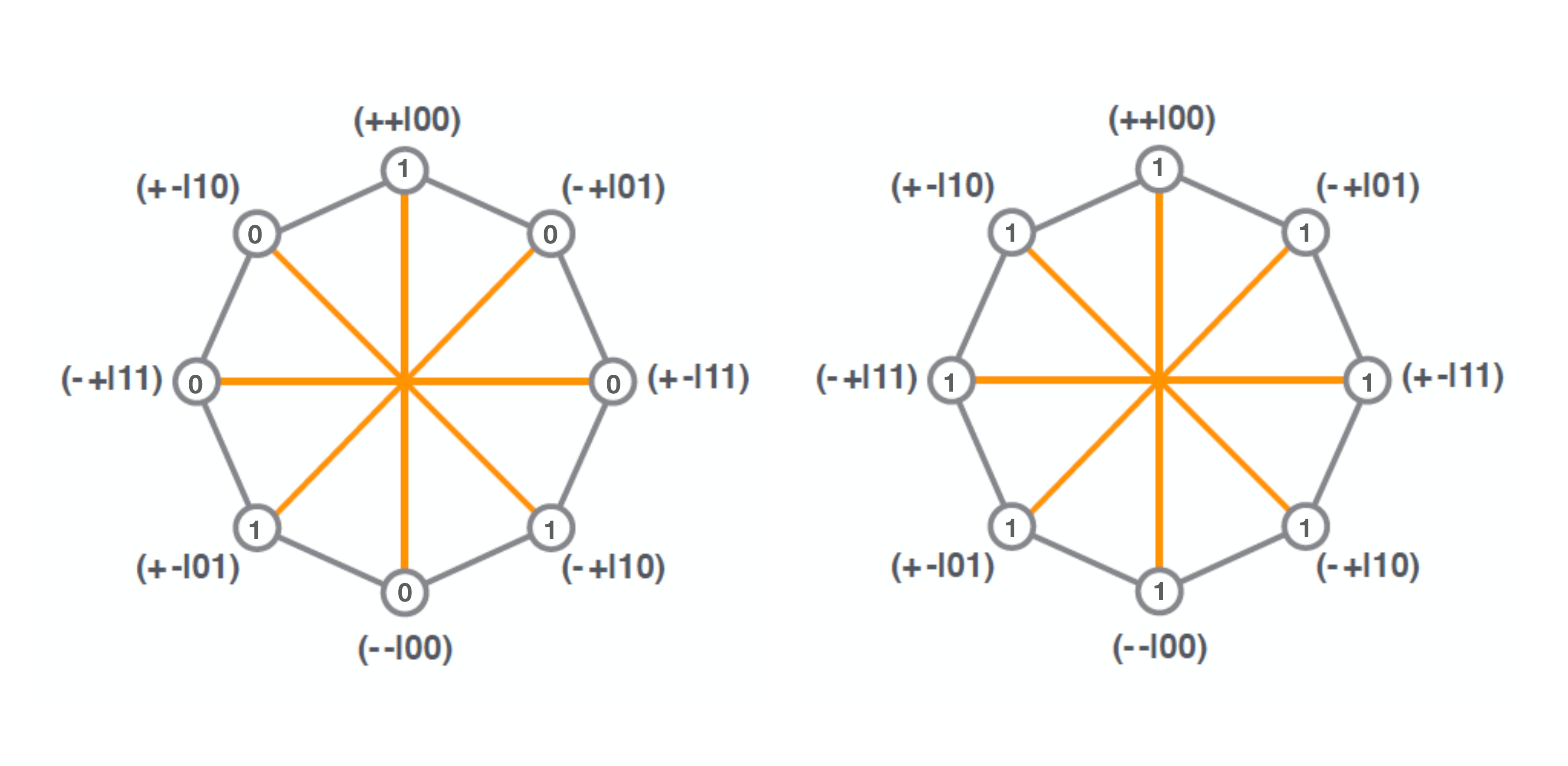}
\vspace{-10mm}
\caption{Examples of photon-assignments to events in the CHSH exclusivity graph. Left -- one photon, right -- two photons.}
\label{fig4}
\end{figure}

%%%%%%%%%%%%%%%%%%%%%%%%%%%%%%%%

Now, consider the same CHSH scenario, but this time inject two indistinguishable photons to the system. The exclusivity and particle number conservation imply that there could be at most two photons per context. The possible occupation numbers are 0, 1 or 2. Since each context consists of only two events, one can assign a single photon to each event, see Fig. \ref{fig4}. Therefore, for two photons ${\cal C}=8$, which is the maximal possible sum of non-contextually assigned occupation numbers over all events. We see that the bound depends on the number of particles. We divide (\ref{CHSH3}) by $N$ and rewrite it as
\begin{eqnarray}
\frac{ \langle n(+-|11)\rangle}{N} +\frac{\langle n(-+|11)\rangle}{N} +\frac{\langle n(+-|01)\rangle}{N} &+& \nonumber \\ \frac{\langle n(-+|01)\rangle}{N} + \frac{\langle n(+-|10)\rangle}{N} +\frac{\langle n(-+|10)\rangle}{N} &+& \label{CHSH4} \\ \frac{\langle n(++|00)\rangle}{N} +\frac{\langle n(--|00) \rangle}{N} &\leq& \mathcal P_C(N),  \nonumber
\end{eqnarray}
where $\mathcal P_C(N)={\cal C}/N$. For a single photon $\mathcal P_C(1)=3$, whereas for two photons $\mathcal P_C(2)=4$. Therefore, for $N=2$ there is no violation and the measurements can be described by NC occupation number assignments.
 
As far as we know, the value $\mathcal P_C(2)=4$ cannot be reached in any experimental setup, although it is allowed in our model. This is becasue the maximal quantum value of $2+\sqrt{2}$ (attainable in the CHSH scenario) does not depend on the physical implementation of the experiment. In particular it does not depend on the dimension of the state space of the physical system, which in our case translates to independence on the particle number $N$.

Finally, let us consider the classical limit $\langle N \rangle \gg 1$. This time the system is described by a classical light beam of intensity $I$ for which the inequality (\ref{CHSH4}) reads
\begin{eqnarray}
\frac{I(+-|11)}{I} +\frac{I(-+|11)}{I} +\frac{I(+-|01)}{I} &+& \nonumber \\ \frac{I(-+|01)}{I} + \frac{I(+-|10)}{I} +\frac{I(-+|10)}{I} &+& \label{CHSHI} \\ \frac{I(++|00)}{I} +\frac{I(--|00) }{I} &\leq& \mathcal P_{cl},  \nonumber
\end{eqnarray}
where $N$ in the denominator of (\ref{CHSH4}) was replaced by $\langle N\rangle$ due to the particle number uncertainty.
The maximal experimentally attainable value of the left-hand side is still $2+\sqrt{2}$ becasue the classical beam in any linear optical setup behaves in the same way as a single-photon probability amplitude.
The right-hand side can be evaluated in two ways. Firstly, in the classical limit the total intensity $I$ that is distributed between the events can be treated as a continuous property. Therefore, in the NC model one can assign $I/2$ to each event and as a result $\mathcal P_{cl}=4$. This is the main result in this section: The corresponding CHSH inequality (\ref{CHSHI}) cannot be violated by classical light. 

\begin{figure}[t]
\includegraphics[width=0.5 \textwidth,trim=4 4 4 4,clip]{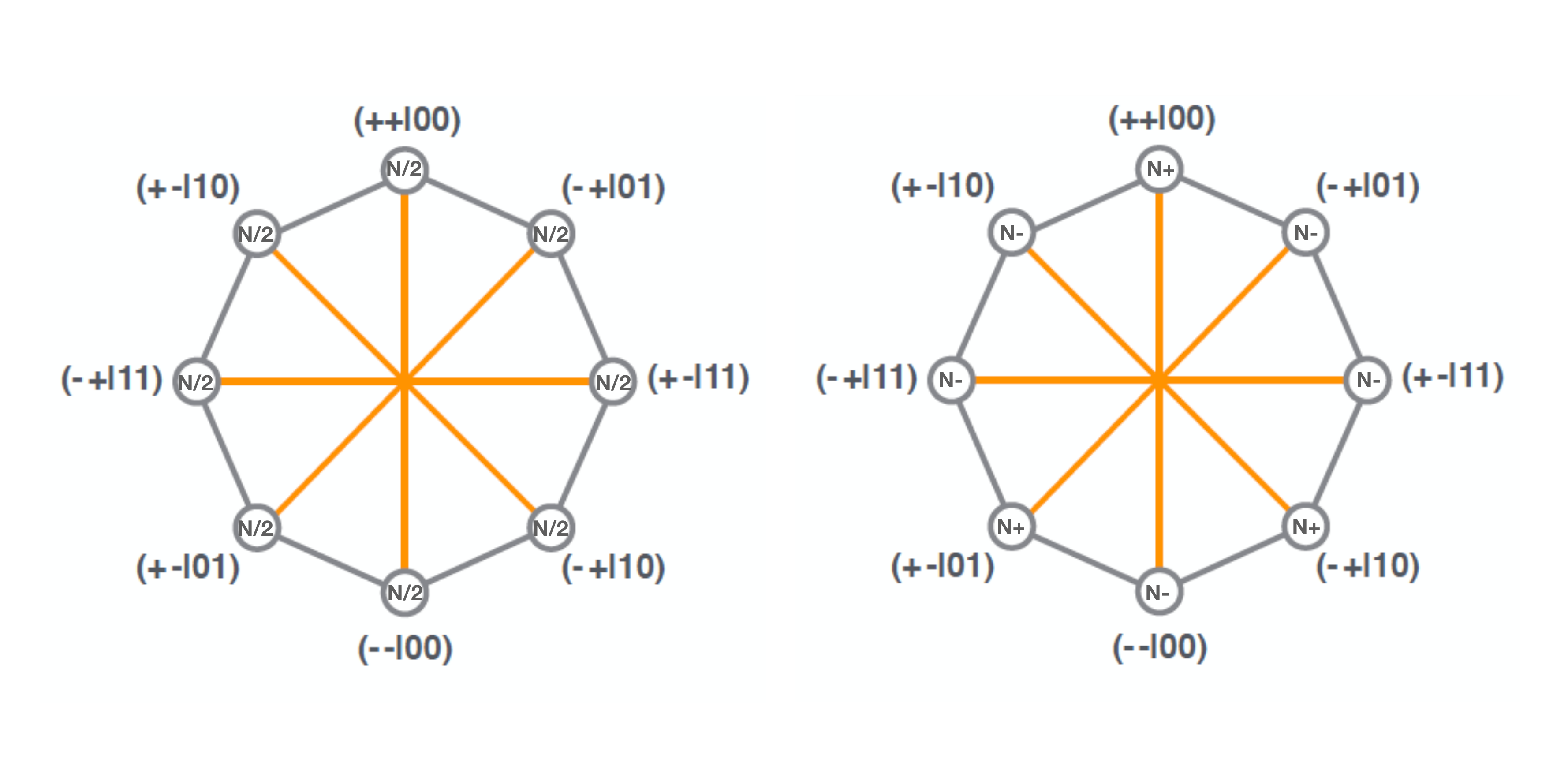}
\vspace{-10mm}
\caption{Assignment of photons to events in the exclusivity graph of the CHSH scenario. Left -- even number of photons, right -- odd number of photons, where $N_+=(N+1)/2$ and $N_-=(N-1)/2$}
\label{fig5}
\end{figure}

The other approach, which also confirms the above result, does not assume that the intensity is a continuous property. We consider two cases. First, let us take even $N$. The number of photons per context cannot be greater than $N$ and since each context contains two events, one simply assigns $N/2$ photons per context. This leads to $\mathcal P_C(N)=4$. Next, we consider odd $N$. In this case it is easy to show that one can assign $(N-1)/2$ photons to five events and $(N+1)/2$ photons to three events, such that there are at most $N$ photons per each context, see Fig. \ref{fig5}. As a result one gets $\mathcal P_C(N)=4-\frac{1}{N}$. In the classical limit $N$ is undetermined, therefore $\mathcal P_{cl} = \langle \mathcal P_C(N) \rangle$. However, since $\langle N \rangle \gg 1$ the dominating terms in $\langle \mathcal P_C(N) \rangle$ correspond to large values of $N$ and hence $\mathcal P_{cl} \approx 4$.

%%%%%%%%%%%%%%%%%%%%%%%%%%%%%%%%

%%%%%%%%%%%%%%%%%%%%%%%%%%%%%%%%

To conclude, we see that in the experiments discussed above light beams, as expected, do not exhibit any quantum behaviour. Quantum behaviour is only possible for a single photon and already for $N>1$ one observes noncontextual (classical) behaviour. This is because for $N\geq2$ the non-contextual bound $\mathcal P_C(N)$ is either 4 or $4-\frac{1}{N}$ and is always greater than the physically attainable value of $2+\sqrt{2} \approx 3.41$. We have only considered the CHSH scenario, however in the Methods section we show that for an arbitrary contextuality scenario the bound $\mathcal P_{cl}$ is always greater or equal than what can be achieved in the classical limit and therefore classical systems are always noncontextual and can never violate any Bell-type inequality. 

%We introduce a model for Bell-type tests on classical waves, which directly represents the ideas in Refs.  Cabello \cite{CabelloSimulations} and Spreeuw \cite{Spreeuw01}.
%In this model the \emph{relative} intensity of light in each mode represents a probability of a joint detection from the Bell scenario - namely $p(v_i)$ for some vertex $v_i$ of the exclusivity graph. Due to the conservation of energy in the multiport the normalization of probabilities $\sum_i p(v_i)\leq 1$ is fulfilled for each context $\mathcal C_k$. The normalization is always strict for the physical multiport, however it need not be strict from the point of view of a Bell inequality, since some joint probabilities might not directly appear in the inequality (for example in the CHSH and Mermin case). In this model however the exclusivity of events is not present - the wave is divided into beams which are present in any mode for which $p(v_i)>0$. We may observe how the exclusivity is lost when starting from a single photon model - this corresponds to the contextual version of the Bell inequalities on a multiport treated as a single system. If single photons are emitted and detected, the exclusivity is present. It is lost for the appropriately strong beam. This suggests, that the classical limiting case should not be treated as the model with continuous output. Actually for fixed settings the intensities in each modes are deterministic - they are always equal to quantum probabilities they simulate.

%%%%%%%%%%%%%%%%%%%%%%%%%%%%%%%%%%%%%%%%%%%%%%%%%%%%%%%%%%%

\section{Discussion}

It is commonly accepted that the violation of some Bell inequality is an indicator of non-classical behaviour. It turns out that these violations are strongly related to the wave-particle duality. In quantum theory the same object can manifest a wave-like behaviour in one experiment and a particle-like behaviour in some other experiment. The wave-particle duality does not occur in classical theory, i.e., classical objects are either waves or particles, but never both. No one ever doubts that classical particles are localized and discrete whereas classical waves are delocalized and continuous.

Bell inequalities are derived under assumption that measured properties, such as position, are well defined and the phenomenon of superposition does not occur. This is a typical particle-like approach which can be partially justified in quantum regime by the fact that at the end of each experiment one registers a single click of some detector. However, the violation of the corresponding inequality implies that this assumption needs to be reconsidered. This is because the measurements in the Bell tests exploit the phenomenon of superposition and in this sense the violation is a manifestation of the wave-like behaviour. 

The situation is different in the classical theory in which there is no wave-particle duality. In particular, the classical light behaves as a wave all the time and instead of clicks one registers continuous intensities. Therefore, there is no justification to apply the standard Bell inequality to such system. In this case Bell inequalities need to be rederived using properly chosen assumptions. This is what we show in this work. 

Our results have the following implications on the previously discussed Bell-type tests with classical light. It was stressed by Snoke \cite{Snoke14} and Qian \emph{et.al.} \cite{Eberly2015} that violation of some Bell inequality by classical light is in fact a confirmation of the presence of some kind of entanglement. Here we show that the value of $2+\sqrt{2}$ obtainable by classical light can indicate some form of strong correlations, but these correlations are fully describable by a noncontextual model. Therefore, classical entanglement does not give rise to contextual behaviour, it does not have any features of quantum entanglement apart from mathematical analogy on the level of complex vector spaces.    

Next, Frustaglia \emph{et. al.} \cite{CabelloSimulations} argued that the bounds on the strength of quantum correlations (so called \emph{Tsirelson bounds}) are not restricted to quantum theory but arise naturally in classical systems that simulate quantum correlations. There are attempts to derive these bounds using only the exclusivity structure of detection events \cite{Exclusivity,LO}. However, we showed that in the CHSH scenario discussed above the classical bound resulting from the exclusivity structure of detection events is $\mathcal P_{cl}=4$. Therefore, the value of $2+\sqrt{2}$ in the classical regime must originate from some physical constraints. 

It is interesting to understand why quantum systems and classical light lead to the same value. The reason for this is that quantum probability amplitudes and classical electromagnetic amplitudes transform in the same way (apart from the nondeterministic quantum collapse induced by the measurement). This is why classical wave theory can simulate some aspects of quantum theory. Nevertheless, the interpretation of both amplitudes is completely different. The value of $2+\sqrt{2}$ would be much more fundamental if it resulted from both, the exclusivity structure of events and from the transformations allowed by the theory. 

Finally, we would like to note that our work leads to an open problem. Is it possible to find a classical system, other than a classical wave, which would provide a value larger than $2+\sqrt{2}$? Perhaps there are some additional constraints preventing this from happening.

%%%%%%%%%%%%%%%%%%%%%%%%%%%%%%%%%%%%%%%%%%%%%%%%%%%%%%%%

\section{Methods}

\subsection{Nonlocality and contextuality}

Original Bell inequalities are statistical tests that verify whether the statistics of measurements' outcomes performed on spatially separated systems fulfill the conditions of locality and \emph{realism} \cite{CHSH}. Locality means that any action executed on one system does not affect the other one. Realism means that each measurable property has a well defined outcome, irrespective of whether this property is measured or not. Initially Bell-type tests were derived from the properties of probability distributions for correlations of measurement outcomes \cite{CHSH}. It was early recognised by Fine \cite{Fine82} that any Bell inequality is equivalent to the existence of a joint probability distribution (JPD) for all measurable properties. 

Violation of Bell inequalities by quantum systems, typically referred to as \emph{quantum nonlocality}, is an example of a broader class of phenomena -- \emph{quantum contextuality} \cite{KS}. Simply speaking, contextuality is a property of a physical system where the outcome of some property $A$ may depend on whether it is measured with $B$ or with $C$. In typical contextuality scenario one considers a system with a number of measurable properties. The goal is to find a noncontextual (NC) assignment of outcomes to all measurements, i.e., to assign an outcome to each property in a way that does not depend on what is assigned to other properties. Just like in the case of nonlocality, contextuality is equivalent to a lack of JPD for all measurable properties \cite{Fine82}. Finally, note that noncontextual scenarios do not require space-like separated measurements, therefore Bell-type scenarios constitute a subclass of contextual scenarios, in which the commeasurable observables can be identified with spatially separated physical systems \cite{Markiewicz14}. This is the reason why we focus on a more general phenomenon of contextuality.

Contextuality scenarios underline the role of the exclusivity structure of events behind Bell-type inequalities, which is not explicit in the standard correlation-based approach. In a series of papers \cite{CSW, CabelloBasicGraphs, Exclusivity, CabelloMermin} Cabello \emph{et. al.} show that all contextuality tests can be derived from the exclusivity structure of measurement events. Assume that the set of all events $v_i$, the precise meaning of which is defined separately in the physical scenario, can be decomposed into subsets $\mathcal C_k=\{v_r\}$ called \emph{measurement contexts}, such that all the events in a single context correspond to outcomes of some experiment. Therefore, the events in a single context can be jointly measured. Because of various constraints, some of the events cannot be simultaneously true, the property which is known as \emph{exclusivity} in probability theory. We point out that the notion of exclusivity is a fundamental property of Kolmogorovian probability theory, since elementary events, which constitute a sample space in any statistical model, must be mutually exclusive. 

The exclusivity structure of the set of events for a contextuality test can be represented in the \emph{exclusivity graph} \cite{CSW} in which adjacent vertices represent exclusive events. An example of such a graph is presented in Fig. \ref{fig3} c). For commeasurable observables $A_i$ with outcomes $a_i$, each vertex represents an event, which is a conjunction of all single detection events for a fixed experimental context. Namely $v_k=(a_1,\ldots,a_n|A_1,\ldots,A_n)$, where (with a slight abuse of notation) the set $\{a_i\}$ represents actual measurement outcomes, obtained for \emph{fixed} properties $\{A_i\}$ that constitute a measurement context. 

Two events $v_a=(a_1,\ldots,a_n|A_1,\ldots,A_n)$ and $v_b=(b_1,\ldots,b_n|B_1,\ldots,B_n)$ in some contextuality scenario are exclusive if and only if:
\be
&&(A_1=B_1\,\&\,a_1\neq b_1)\textrm{ OR }(A_2=B_2\,\&\,a_2\neq b_2)\textrm{ OR }\ldots\nonumber\\
&&\textrm{OR }(A_n=B_n\,\&\,a_n\neq b_n).
\label{BellExcl}
\ee
Having defined the exclusivity structure for a given scenario, one may attach different kinds of probabilistic structures to the set of events. We stress that the phrase \emph{probabilistic structure} does not refer to a Kolmogorovian model, instead it can be understood within the context of generalized probabilistic theories \cite{Short10}. The models can be ordered with respect to the strength of allowed correlations and this strength can be measured by the maximal value of $\mathcal P=\sum_i p(v_i)$ allowed within the model. 

We discuss three classes of models. The most restrictive one is the classical Kolmogorovian model. It assumes that all events can be attributed to a single sample space (not necessarily as atomic events, but compound events as well), and that a joint probability distribution exists for all $v_i$. Then the maximal value of ${\mathcal P}$, denoted by ${\mathcal P}_C$, known as \emph{noncontextual local hidden variable} (NCHV) or simply classical bound, is given by the \emph{vertex independence number} of the exclusivity graph -- the maximal number of non-adjacent vertices. To rephrase, the classical bound ${\mathcal P}_C$ corresponds to the maximal number of events that can be assigned truth value $1$. This mathematical model represents a physical situation where all observables are commeasurable. 

The second model, involving stronger correlations, is the quantum probabilistic model \cite{Streater00}. Here each event corresponds to a projector $P(v_i)$. Each projector is assigned a probability $p(v_i)=\Tr(\rho P_i(v_i))$, where $\rho$ represents a quantum state. For the context $\mathcal C_k$ consisting of mutually exclusive events the corresponding projectors make a mutually orthogonal set and therefore:
\be
\sum_{v_i\in\mathcal C_k} P(v_i)\leq \openone.
\label{quantmodel}
\ee
This means that either (equality) the projectors within a context represent a von Neumann measurement or (sharp inequality) a von Neumann measurement extension is possible. The maximum value ${\mathcal P}_Q$  of $\mathcal P$ for a quantum model, known as the \emph{Tsirelson's bound}, is bounded by the Lov\'asz number of the exclusivity graph \cite{CSW}. This bound is often tight \cite{CabelloBasicGraphs}. The quantum model is \emph{contextual} if and only if ${\mathcal P}_C < \sum_i p(v_i) \leq {\mathcal P}_Q$. This means that an assignment of outcome probabilities can be done within each context separately and that the model cannot be extended to a single Kolmogorovian probabilistic model.

The third model, leading to the strongest correlations, is defined by a sole demand that its probabilistic structure fulfills the \emph{exclusivity (E) principle} in its final version \cite{Exclusivity} (whose other variant, known as \emph{local orthogonality}, was independently developed in \cite{LO}). The E principle states that the sum of probabilities corresponding to a subset of pairwise \emph{exclusive} events is bounded by $1$. This principle is a conjunction of two facts: (1) Cabello's principle called the Specker's principle \cite{CabelloMermin}, which states that the set of pairwise exclusive events is jointly exclusive, and (2) the property of a Kolmogorovian model that the sum of probabilities of jointly exclusive events is at most $1$. Therefore, the probabilistic model obeying the E principle can be defined as the set of numbers $0\leq p(v_i)\leq 1$, fulfilling the normalization condition $\sum_{v_i\in\mathcal C_k}p(v_i)\leq 1$ for each context. 

Note that both, classical Kolmogorovian model and the quantum model, obey the E principle. This does not mean that a model obeying the E principle is classical or quantum. There are models obeying the E principle that are more contextual than quantum theory. This is because the maximal possible value ${\mathcal P}_E$ of $\mathcal P$ for the model obeying the E principle can be greater than ${\mathcal P}_Q$. One can derive it using the properties of an exclusivity graph. The detailed derivation is given in the appendix of \cite{CabelloSimple} and states that the maximal possible value is given by the \emph{fractional packing number} of the graph, defined as:
\be
\max\sum_{i}\omega_i, \textrm{ s.t. } (\forall_i\,0\leq\omega_i\leq 1\textrm{ and } \sum_{i\in\mathcal S}\omega_i\leq 1),
\label{fractional}
\ee
The maximum is taken over all cliques $\mathcal S$ of the graph. A clique is a subset of mutually adjacent vertices (pairwise exclusive events). Therefore the cliques of an exclusivity graph correspond to all the contexts $\mathcal C_k$. The Lov\'asz number is bounded by the fractional packing number. However in many cases the fractional packing number is larger than the Lov\'asz number.

To conclude, the bounds for the three models obey the following relation
\begin{equation}
{\mathcal P}_C \leq {\mathcal P}_Q \leq {\mathcal P}_E.
\end{equation}
For example, in the case of the exclusivity graph presented in Fig. \ref{fig3} c) one has ${\mathcal P}_C = 3$, ${\mathcal P}_Q=2+\sqrt{2}$ and ${\mathcal P}_E=4$. In general, if the sum of probabilities of events in the exclusivity graph is bounded by ${\mathcal P}_C $ then we say that the model is noncontextual. Otherwise it is contextual. 

\subsection{Contextuality and indistinguishability}

Now we discuss events that occur in scenarios in which instead of a single photon one uses $N$ indistinguishable photons. If the number of photons $N$ in the system increases so does the number of possible detection events. This affects the structure of the exclusivity graph and that of bounds  ${\mathcal P}_C $,  ${\mathcal P}_Q$ and  ${\mathcal P}_E$. Instead of changing the graph, one can keep the exclusivity graph in the same form and change the range of values that are assigned to each event, which warrants a slight redefinition of the entire scenario. Here we focus on the classical Kolmogorovian model because we want to find a criterion for which the system of $N$ indistinguishable photons is noncontextual. We will show that within the Kolmogorovian model the bound $\mathcal P_C$ depends on $N$, i.e., ${\mathcal P}_C = {\mathcal P}_C (N)$. Finally, we show that in the classical limit ${\mathcal P}_{cl}={\mathcal P}_E$, therefore classical light, for which the ratio of output intensities to the initial intensity is the same as the probabilities generated by a single photon, is always noncontextual.  

For $N=1$ one assigns truth values $1/0$ to each event and tries to maximise the number of events that are assigned $1$ under the exclusivity constraint. The value $1$ corresponds to events that will be observed in an experiment, provided a proper measurement is performed, whereas $0$ corresponds to events that will not be observed. However, $1$ can be interpreted as assignment of a photon to an event and $0$ as an assignment of the absence of a photon. One can generalise this approach to the case $N>1$. This time each event $v_i$ is assigned a value $n_i=0,1,\dots,N$. The values $n_i$ can be interpreted as occupation numbers. This approach was suggested in \cite{Kurzynski2017}. The exclusivity of $v_i$ and $v_j$ translates to $n_i+n_j\leq N$. In general, the sum of occupation numbers in each context $\sum_{v_i\in\mathcal C_k} n_i \leq N$. Therefore, the exclusivity of events implies that the number of particles per context cannot be larger than the total number of particles. If the context consists of all measurement outcome events then the number of particles in it must be equal to $N$. This corresponds to the particle-number conservation. In full analogy to the original model, which assigns probabilities to the events, the strength of correlations for assigning particles to the events is represented by the expression $\mathcal P(N)=\frac{1}{N} \sum_{i}  n_i$, where the sum is taken over all events in the exclusivity graph. The Kolmogorovian model in the original scenario translates to an occupation-based model with fixed global assignment of occupation numbers.

We are looking for the maximal value $\mathcal P_C(N)$ of the expression $\mathcal{P}(N)$ optimized over all allowed NC global assignments. For $N=1$ the NC occupation-based model has the same constraints as the deterministic model with assigned probabilities, thus $\mathcal P_C(1)=\mathcal P_C$. However, for $N>1$ the constraints are different and $\mathcal P_C(N) \neq \mathcal P_C$. In particular, we are interested in the classical limit $\langle N \rangle \gg 1$. The system, corresponding to a classical light beam, is modelled by a collection of photons whose total number is undetermined, but its average number is large. We are looking for $\mathcal P_{cl}$ which is the maximum of $\frac{1}{\langle N \rangle} \sum_{i}  \langle n_i \rangle$. We write $I_i \equiv \langle n_i \rangle$. We call $I_i$ the intensity assigned to the event $v_i$. There is a fundamental difference between intensities and occupation numbers. In general $\langle n_i \rangle \neq n_i$, however classical limit is deterministic and therefore $\langle I_i \rangle = I_i$. In reality $I_i$ fluctuates as $\sqrt{\langle N \rangle}$, but we will come back to this problem in a moment. Moreover, unlike occupation numbers, intensities can be treated as continuous variables $I_i \in [0,I]$, where $I$ is the total intensity. In this case we are looking for $\mathcal P_{cl}$ which is a maximum of 
\begin{equation}
\label{limcase}
\sum_{i} \frac{I_i}{I},
\end{equation}
where $I=\langle N \rangle$ is the total intensity. The ratios $0 \leq I_i/I \leq 1$ are continuous numbers and the only constraint on them is that within each context:
$\sum_{v_i\in\mathcal C_k} \frac{I_i}{I} \leq 1$.
As we mentioned above $I_i$ is not truly deterministic, but the ratio $I_i/I$ fluctuates as $1/\sqrt{\langle N \rangle}$, therefore in the classical limit $\langle N \rangle \gg 1$ these fluctuations are of no importance to us.

By taking $p_i=I_i/I$ the above model is equivalent to the third model of the original scenario, which only needs to obey the E principle. Therefore the noncontextuality bound in the classical limit is $\mathcal P_{cl}=\mathcal P_E$. Since the classical light when passing a linear optical setup evolves in the same way as the single-particle quantum amplitudes, we have:
\begin{equation}
\sum_{i} \frac{I_i}{I}\leq \mathcal P_Q \leq \mathcal P_E = \mathcal P_{cl},
\end{equation}
which proves that classical light is always noncontextual.

%The above analysis explicitly shows the source of confusion about the possible non-classical nature of correlations arising in experiments with classical waves. Namely the transition from a single-particle model to more particles radically changes the exclusivity of detection events. In the single-particle case this exclusivity is equivalent to the probabilistic notion of exclusivity of single detection clicks within an experimental context, whereas for many particles it transforms into the conservation of particle number within the context. In the limiting case the particle conservation transforms into the conservation of energy of an electromagnetic wave. These transformations change the logical structure of the experiment. For a single particle detection events are maximally exclusive, for more particles some events cease to be exclusive, and in the limiting case the exclusivity of detection events is completely lost. This incremental decrease of exclusivity in the transition from a single photon to a macroscopic wave increases the value of the maximal attainable strength of correlations in a given Bell scenario for a non-contextual value assignment. We showed that in the limiting case this increased value is always larger or equal than the value which is attainable in experiments with macroscopic waves.

\subsection{Non-applicability of the methods used to derive the quantum bound $\mathcal P_Q$ with the E principle}

In \cite{Exclusivity} a general method of derivation of the maximal quantum bound for a given Bell-type scenario was introduced. This method is based on the idea, that the E principle can be applied not only to a single copy of a Bell-type test represented by some exclusivity graph $G$, but also to $k$ independent runs of such identical Bell tests carried on different physical systems. Such a compound experiment is described by an exclusivity graph that is represented by the so called disjunctive product $G^k$ of $k$ copies of the graph $G$ \cite{ORP}. It is assumed that the E principle should be applicable also to any clique (representing measurement context) in the graph $G^k$. It might seem that such a procedure should not lead to any new conclusions, since the carried $k$ experiments are completely independent. However it turns out that application of the E principle to a clique in $G^k$ restricts the possible assignments in much stronger way than in the case when it is applied to a single copy of $G$. This is because each vertex $V$ of $G^k$ represents a joint event, which consists of a conjunction of $k$ independent events corresponding to some $k$ vertices $\{v_i\}$ of $G$. Therefore the assigned probability factorizes into $p(V)=p(v_1)\cdot\ldots\cdot p(v_k)$. It can be easily shown that application of E principle to some clique in $G^k$ consisting of some number of vertices $V$ places stronger restriction on the possible values of $p(v_i)$. 

The easiest possible case is given by the contextuality test of Klychko-Can-Binicioglu-Schumovsky (KCBS) \cite{KCBS}, in which the original exclusivity graph $G$ is a 5-cycle, and therefore the E principle allows for assignment of the probability at most $\tfrac{1}{2}$ to each vertex leading to the bound $\mathcal P_E=\tfrac{5}{2}$. On the other hand the graph $G^2$ representing the exclusivity structure for two independent KCBS experiments has a $5$-vertex clique $K$, and the E principle applied to $K$ allows for assignment of a probability at most $\tfrac{1}{5}$ to each vertex of $K$. Note that a probability assigned to each vertex $V_i$ of $K$ is a product of probabilities corresponding to different subsets of vertices of $G$: $p(V_i)=p(v_i)p(u_i)$. Now, $K$ can be chosen such that the vertices in $K$, which are of the form $\{v_i \times u_j\}$, contain all the vertices from the elementary graphs $G$. Hence assignments $p(v_i)$ and $p(u_j)$ are also assignments to the entire graph $G$ of a single KCBS test. Since $\sum_{i=1}^5 p(v_i)p(u_i)\leq 1$, and the set $\{p(v_i)\}$ is any permutation of the set $\{p(u_j)\}$, we obtain the restriction $\sum_i p(v_i)\leq \sqrt{5}$, which exactly reproduces the quantum bound $\mathcal P_Q$. 

Now it can be easily seen that the above derivation cannot be applied to the case of classical waves, described by the NC occupation-based model defined in the previous section. This is because in the limiting case of the modified model the vertices of the graph are assigned relative intensities of light instead of probabilities of events \eqref{limcase}. Taking two such independent experiments one cannot meaningfully create a product graph, the vertices of which would correspond to relative intensities which are products of two intensities from the single experiments. This follows from the fact that the intensity of two independent waves is not a product of their intensities but instead it is their sum (assuming they propagate in a linear medium). Instead, the joint probability of two independent events is a product of their probabilities. This implies that the entire derivation of the bound cannot be performed in the case of correlations for classical waves. Nevertheless classical waves still obey the bound $\mathcal P_Q$. As we showed, this cannot be directly derived from the exclusivity structure of the experiment. It holds because classical waves follow the same evolution rules as quantum probability amplitudes.

\section{Acknowledgements}

M.M. and P.K. were supported by the National Science Centre in Poland through NCN Grant No. 2014/14/E/ST2/00585. D.K. was supported by the National Research Foundation and Ministry of Education in Singapore.

\bibliographystyle{apsrev4-1}
%\bibliography{ClEntRev}

\begin{thebibliography}{36}%
\makeatletter
\providecommand \@ifxundefined [1]{%
 \@ifx{#1\undefined}
}%
\providecommand \@ifnum [1]{%
 \ifnum #1\expandafter \@firstoftwo
 \else \expandafter \@secondoftwo
 \fi
}%
\providecommand \@ifx [1]{%
 \ifx #1\expandafter \@firstoftwo
 \else \expandafter \@secondoftwo
 \fi
}%
\providecommand \natexlab [1]{#1}%
\providecommand \enquote  [1]{``#1''}%
\providecommand \bibnamefont  [1]{#1}%
\providecommand \bibfnamefont [1]{#1}%
\providecommand \citenamefont [1]{#1}%
\providecommand \href@noop [0]{\@secondoftwo}%
\providecommand \href [0]{\begingroup \@sanitize@url \@href}%
\providecommand \@href[1]{\@@startlink{#1}\@@href}%
\providecommand \@@href[1]{\endgroup#1\@@endlink}%
\providecommand \@sanitize@url [0]{\catcode `\\12\catcode `\$12\catcode
  `\&12\catcode `\#12\catcode `\^12\catcode `\_12\catcode `\%12\relax}%
\providecommand \@@startlink[1]{}%
\providecommand \@@endlink[0]{}%
\providecommand \url  [0]{\begingroup\@sanitize@url \@url }%
\providecommand \@url [1]{\endgroup\@href {#1}{\urlprefix }}%
\providecommand \urlprefix  [0]{URL }%
\providecommand \Eprint [0]{\href }%
\providecommand \doibase [0]{http://dx.doi.org/}%
\providecommand \selectlanguage [0]{\@gobble}%
\providecommand \bibinfo  [0]{\@secondoftwo}%
\providecommand \bibfield  [0]{\@secondoftwo}%
\providecommand \translation [1]{[#1]}%
\providecommand \BibitemOpen [0]{}%
\providecommand \bibitemStop [0]{}%
\providecommand \bibitemNoStop [0]{.\EOS\space}%
\providecommand \EOS [0]{\spacefactor3000\relax}%
\providecommand \BibitemShut  [1]{\csname bibitem#1\endcsname}%
\let\auto@bib@innerbib\@empty
%</preamble>
\bibitem [{\citenamefont {Pan}\ \emph {et~al.}(2012)\citenamefont {Pan},
  \citenamefont {Chen}, \citenamefont {Lu}, \citenamefont {Weinfurter},
  \citenamefont {Zeilinger},\ and\ \citenamefont {\ifmmode~\dot{Z}\else
  \.{Z}\fi{}ukowski}}]{PanReview}%
  \BibitemOpen
  \bibfield  {author} {\bibinfo {author} {\bibfnamefont {J.-W.}\ \bibnamefont
  {Pan}}, \bibinfo {author} {\bibfnamefont {Z.-B.}\ \bibnamefont {Chen}},
  \bibinfo {author} {\bibfnamefont {C.-Y.}\ \bibnamefont {Lu}}, \bibinfo
  {author} {\bibfnamefont {H.}~\bibnamefont {Weinfurter}}, \bibinfo {author}
  {\bibfnamefont {A.}~\bibnamefont {Zeilinger}}, \ and\ \bibinfo {author}
  {\bibfnamefont {M.}~\bibnamefont {\ifmmode~\dot{Z}\else \.{Z}\fi{}ukowski}},\
  }\href {\doibase 10.1103/RevModPhys.84.777} {\bibfield  {journal} {\bibinfo
  {journal} {Rev. Mod. Phys.}\ }\textbf {\bibinfo {volume} {84}},\ \bibinfo
  {pages} {777} (\bibinfo {year} {2012})}\BibitemShut {NoStop}%
\bibitem [{\citenamefont {Hong}\ \emph {et~al.}(1987)\citenamefont {Hong},
  \citenamefont {Ou},\ and\ \citenamefont {Mandel}}]{HOM}%
  \BibitemOpen
  \bibfield  {author} {\bibinfo {author} {\bibfnamefont {C.~K.}\ \bibnamefont
  {Hong}}, \bibinfo {author} {\bibfnamefont {Z.~Y.}\ \bibnamefont {Ou}}, \ and\
  \bibinfo {author} {\bibfnamefont {L.}~\bibnamefont {Mandel}},\ }\href
  {\doibase 10.1103/PhysRevLett.59.2044} {\bibfield  {journal} {\bibinfo
  {journal} {Phys. Rev. Lett.}\ }\textbf {\bibinfo {volume} {59}},\ \bibinfo
  {pages} {2044} (\bibinfo {year} {1987})}\BibitemShut {NoStop}%
\bibitem [{\citenamefont {Lo}\ \emph {et~al.}(2016)\citenamefont {Lo},
  \citenamefont {Li}, \citenamefont {Yabushita}, \citenamefont {Chen},
  \citenamefont {Luo},\ and\ \citenamefont {Kobayashi}}]{LO16}%
  \BibitemOpen
  \bibfield  {author} {\bibinfo {author} {\bibfnamefont {H.-P.}\ \bibnamefont
  {Lo}}, \bibinfo {author} {\bibfnamefont {C.-M.}\ \bibnamefont {Li}}, \bibinfo
  {author} {\bibfnamefont {A.}~\bibnamefont {Yabushita}}, \bibinfo {author}
  {\bibfnamefont {Y.-N.}\ \bibnamefont {Chen}}, \bibinfo {author}
  {\bibfnamefont {C.-W.}\ \bibnamefont {Luo}}, \ and\ \bibinfo {author}
  {\bibfnamefont {T.}~\bibnamefont {Kobayashi}},\ }\href
  {http://dx.doi.org/10.1038/srep22088} {\bibfield  {journal} {\bibinfo
  {journal} {Scientific Reports}\ }\textbf {\bibinfo {volume} {6}},\ \bibinfo
  {pages} {22088 EP } (\bibinfo {year} {2016})},\ \bibinfo {note}
  {article}\BibitemShut {NoStop}%
\bibitem [{\citenamefont {Jin}\ \emph {et~al.}(2017)\citenamefont {Jin},
  \citenamefont {Chen}, \citenamefont {Jing}, \citenamefont {Ren},
  \citenamefont {Zhao}, \citenamefont {Shimizu},\ and\ \citenamefont
  {Lu}}]{Jin17}%
  \BibitemOpen
  \bibfield  {author} {\bibinfo {author} {\bibfnamefont {R.-B.}\ \bibnamefont
  {Jin}}, \bibinfo {author} {\bibfnamefont {G.-Q.}\ \bibnamefont {Chen}},
  \bibinfo {author} {\bibfnamefont {H.}~\bibnamefont {Jing}}, \bibinfo {author}
  {\bibfnamefont {C.}~\bibnamefont {Ren}}, \bibinfo {author} {\bibfnamefont
  {P.}~\bibnamefont {Zhao}}, \bibinfo {author} {\bibfnamefont {R.}~\bibnamefont
  {Shimizu}}, \ and\ \bibinfo {author} {\bibfnamefont {P.-X.}\ \bibnamefont
  {Lu}},\ }\href {\doibase 10.1103/PhysRevA.95.062341} {\bibfield  {journal}
  {\bibinfo  {journal} {Phys. Rev. A}\ }\textbf {\bibinfo {volume} {95}},\
  \bibinfo {pages} {062341} (\bibinfo {year} {2017})}\BibitemShut {NoStop}%
\bibitem [{\citenamefont {Ramanathan}\ \emph {et~al.}(2011)\citenamefont
  {Ramanathan}, \citenamefont {Paterek}, \citenamefont {Kay}, \citenamefont
  {Kurzy\ifmmode~\acute{n}\else \'{n}\fi{}ski},\ and\ \citenamefont
  {Kaszlikowski}}]{Ravi11}%
  \BibitemOpen
  \bibfield  {author} {\bibinfo {author} {\bibfnamefont {R.}~\bibnamefont
  {Ramanathan}}, \bibinfo {author} {\bibfnamefont {T.}~\bibnamefont {Paterek}},
  \bibinfo {author} {\bibfnamefont {A.}~\bibnamefont {Kay}}, \bibinfo {author}
  {\bibfnamefont {P.}~\bibnamefont {Kurzy\ifmmode~\acute{n}\else
  \'{n}\fi{}ski}}, \ and\ \bibinfo {author} {\bibfnamefont {D.}~\bibnamefont
  {Kaszlikowski}},\ }\href {\doibase 10.1103/PhysRevLett.107.060405} {\bibfield
   {journal} {\bibinfo  {journal} {Phys. Rev. Lett.}\ }\textbf {\bibinfo
  {volume} {107}},\ \bibinfo {pages} {060405} (\bibinfo {year}
  {2011})}\BibitemShut {NoStop}%
\bibitem [{\citenamefont {Aspden}\ \emph {et~al.}(2016)\citenamefont {Aspden},
  \citenamefont {Padgett},\ and\ \citenamefont {Spalding}}]{SingPhoton}%
  \BibitemOpen
  \bibfield  {author} {\bibinfo {author} {\bibfnamefont {R.~S.}\ \bibnamefont
  {Aspden}}, \bibinfo {author} {\bibfnamefont {M.~J.}\ \bibnamefont {Padgett}},
  \ and\ \bibinfo {author} {\bibfnamefont {G.~C.}\ \bibnamefont {Spalding}},\
  }\href {\doibase 10.1119/1.4955173} {\bibfield  {journal} {\bibinfo
  {journal} {American Journal of Physics}\ }\textbf {\bibinfo {volume} {84}},\
  \bibinfo {pages} {671} (\bibinfo {year} {2016})},\ \Eprint
  {http://arxiv.org/abs/https://doi.org/10.1119/1.4955173}
  {https://doi.org/10.1119/1.4955173} \BibitemShut {NoStop}%
\bibitem [{\citenamefont {Suppes}\ \emph {et~al.}(1996)\citenamefont {Suppes},
  \citenamefont {de~Barros},\ and\ \citenamefont {Sant'Anna}}]{Suppes96}%
  \BibitemOpen
  \bibfield  {author} {\bibinfo {author} {\bibfnamefont {P.}~\bibnamefont
  {Suppes}}, \bibinfo {author} {\bibfnamefont {J.~A.}\ \bibnamefont
  {de~Barros}}, \ and\ \bibinfo {author} {\bibfnamefont {A.~S.}\ \bibnamefont
  {Sant'Anna}},\ }\href {https://arxiv.org/abs/quant-ph/9606019} {\bibfield
  {journal} {\bibinfo  {journal} {arXiv:quant-ph/9606019}\ } (\bibinfo {year}
  {1996})}\BibitemShut {NoStop}%
\bibitem [{\citenamefont {Spreeuw}(1998)}]{Spreeuw98}%
  \BibitemOpen
  \bibfield  {author} {\bibinfo {author} {\bibfnamefont {R.~J.~C.}\
  \bibnamefont {Spreeuw}},\ }\href {\doibase doi:10.1023/A:1018703709245}
  {\bibfield  {journal} {\bibinfo  {journal} {Found. Phys.}\ }\textbf {\bibinfo
  {volume} {28}},\ \bibinfo {pages} {361} (\bibinfo {year} {1998})}\BibitemShut
  {NoStop}%
\bibitem [{\citenamefont {Clauser}\ \emph {et~al.}(1969)\citenamefont
  {Clauser}, \citenamefont {Horne}, \citenamefont {Shimony},\ and\
  \citenamefont {Holt}}]{CHSH}%
  \BibitemOpen
  \bibfield  {author} {\bibinfo {author} {\bibfnamefont {J.~F.}\ \bibnamefont
  {Clauser}}, \bibinfo {author} {\bibfnamefont {M.~A.}\ \bibnamefont {Horne}},
  \bibinfo {author} {\bibfnamefont {A.}~\bibnamefont {Shimony}}, \ and\
  \bibinfo {author} {\bibfnamefont {R.~A.}\ \bibnamefont {Holt}},\ }\href
  {\doibase 10.1103/PhysRevLett.23.880} {\bibfield  {journal} {\bibinfo
  {journal} {Phys. Rev. Lett.}\ }\textbf {\bibinfo {volume} {23}},\ \bibinfo
  {pages} {880} (\bibinfo {year} {1969})}\BibitemShut {NoStop}%
\bibitem [{\citenamefont {Spreeuw}(2001)}]{Spreeuw01}%
  \BibitemOpen
  \bibfield  {author} {\bibinfo {author} {\bibfnamefont {R.~J.~C.}\
  \bibnamefont {Spreeuw}},\ }\href {\doibase 10.1103/PhysRevA.63.062302}
  {\bibfield  {journal} {\bibinfo  {journal} {Phys. Rev. A}\ }\textbf {\bibinfo
  {volume} {63}},\ \bibinfo {pages} {062302} (\bibinfo {year}
  {2001})}\BibitemShut {NoStop}%
\bibitem [{\citenamefont {Brunner}\ \emph {et~al.}(2014)\citenamefont
  {Brunner}, \citenamefont {Cavalcanti}, \citenamefont {Pironio}, \citenamefont
  {Scarani},\ and\ \citenamefont {Wehner}}]{Brunner14}%
  \BibitemOpen
  \bibfield  {author} {\bibinfo {author} {\bibfnamefont {N.}~\bibnamefont
  {Brunner}}, \bibinfo {author} {\bibfnamefont {D.}~\bibnamefont {Cavalcanti}},
  \bibinfo {author} {\bibfnamefont {S.}~\bibnamefont {Pironio}}, \bibinfo
  {author} {\bibfnamefont {V.}~\bibnamefont {Scarani}}, \ and\ \bibinfo
  {author} {\bibfnamefont {S.}~\bibnamefont {Wehner}},\ }\href {\doibase
  10.1103/RevModPhys.86.419} {\bibfield  {journal} {\bibinfo  {journal} {Rev.
  Mod. Phys.}\ }\textbf {\bibinfo {volume} {86}},\ \bibinfo {pages} {419}
  (\bibinfo {year} {2014})}\BibitemShut {NoStop}%
\bibitem [{\citenamefont {Qian}\ and\ \citenamefont
  {Eberly}(2011)}]{Eberly2011}%
  \BibitemOpen
  \bibfield  {author} {\bibinfo {author} {\bibfnamefont {X.-F.}\ \bibnamefont
  {Qian}}\ and\ \bibinfo {author} {\bibfnamefont {J.~H.}\ \bibnamefont
  {Eberly}},\ }\href {https://arxiv.org/abs/1011.0693v2} {\bibfield  {journal}
  {\bibinfo  {journal} {Optics Letters}\ }\textbf {\bibinfo {volume} {36}},\
  \bibinfo {pages} {4110} (\bibinfo {year} {2011})}\BibitemShut {NoStop}%
\bibitem [{\citenamefont {Qian}\ and\ \citenamefont
  {Eberly}(2013)}]{Eberly2013}%
  \BibitemOpen
  \bibfield  {author} {\bibinfo {author} {\bibfnamefont {X.-F.}\ \bibnamefont
  {Qian}}\ and\ \bibinfo {author} {\bibfnamefont {J.~H.}\ \bibnamefont
  {Eberly}},\ }\href {https://arxiv.org/abs/1307.3772v1} {\bibfield  {journal}
  {\bibinfo  {journal} {arXiv:1307.3772 [quant-ph]}\ } (\bibinfo {year}
  {2013})}\BibitemShut {NoStop}%
\bibitem [{\citenamefont {Aiello}\ \emph {et~al.}(2015)\citenamefont {Aiello},
  \citenamefont {T\"{o}ppel}, \citenamefont {Marquardt}, \citenamefont
  {Giacobino},\ and\ \citenamefont {Leuchs}}]{Aiello15NJP}%
  \BibitemOpen
  \bibfield  {author} {\bibinfo {author} {\bibfnamefont {A.}~\bibnamefont
  {Aiello}}, \bibinfo {author} {\bibfnamefont {F.}~\bibnamefont {T\"{o}ppel}},
  \bibinfo {author} {\bibfnamefont {C.}~\bibnamefont {Marquardt}}, \bibinfo
  {author} {\bibfnamefont {E.}~\bibnamefont {Giacobino}}, \ and\ \bibinfo
  {author} {\bibfnamefont {G.}~\bibnamefont {Leuchs}},\ }\href {\doibase
  10.1088/1367-2630/17/4/043024} {\bibfield  {journal} {\bibinfo  {journal}
  {New J. Phys.}\ }\textbf {\bibinfo {volume} {17}},\ \bibinfo {pages} {043024}
  (\bibinfo {year} {2015})}\BibitemShut {NoStop}%
\bibitem [{\citenamefont {Wolf}(2007)}]{Wolf07}%
  \BibitemOpen
  \bibfield  {author} {\bibinfo {author} {\bibfnamefont {E.}~\bibnamefont
  {Wolf}},\ }\href@noop {} {\emph {\bibinfo {title} {Introduction to the Theory
  of Coherence and Polarization of Light}}}\ (\bibinfo  {publisher} {Cambridge
  Univ. Press},\ \bibinfo {year} {2007})\BibitemShut {NoStop}%
\bibitem [{\citenamefont {Qian}\ \emph {et~al.}(2015)\citenamefont {Qian},
  \citenamefont {Little}, \citenamefont {Howell},\ and\ \citenamefont
  {Eberly}}]{Eberly2015}%
  \BibitemOpen
  \bibfield  {author} {\bibinfo {author} {\bibfnamefont {X.}~\bibnamefont
  {Qian}}, \bibinfo {author} {\bibfnamefont {B.}~\bibnamefont {Little}},
  \bibinfo {author} {\bibfnamefont {J.}~\bibnamefont {Howell}}, \ and\ \bibinfo
  {author} {\bibfnamefont {J.}~\bibnamefont {Eberly}},\ }\href
  {http://dx.doi.org/10.1364/OPTICA.2.000611} {\bibfield  {journal} {\bibinfo
  {journal} {Optica}\ }\textbf {\bibinfo {volume} {2}},\ \bibinfo {pages} {611}
  (\bibinfo {year} {2015})}\BibitemShut {NoStop}%
\bibitem [{\citenamefont {Snoke}(2014)}]{Snoke14}%
  \BibitemOpen
  \bibfield  {author} {\bibinfo {author} {\bibfnamefont {D.}~\bibnamefont
  {Snoke}},\ }\href {https://arxiv.org/abs/1406.7023} {\bibfield  {journal}
  {\bibinfo  {journal} {arXiv:1406.7023 [quant-ph]}\ } (\bibinfo {year}
  {2014})}\BibitemShut {NoStop}%
\bibitem [{\citenamefont {Frustaglia}\ \emph {et~al.}(2016)\citenamefont
  {Frustaglia}, \citenamefont {Baltan\'as}, \citenamefont
  {Vel\'azquez-Ahumada}, \citenamefont {Fern\'andez-Prieto}, \citenamefont
  {Lujambio}, \citenamefont {Losada}, \citenamefont {Freire},\ and\
  \citenamefont {Cabello}}]{CabelloSimulations}%
  \BibitemOpen
  \bibfield  {author} {\bibinfo {author} {\bibfnamefont {D.}~\bibnamefont
  {Frustaglia}}, \bibinfo {author} {\bibfnamefont {J.~P.}\ \bibnamefont
  {Baltan\'as}}, \bibinfo {author} {\bibfnamefont {M.~C.}\ \bibnamefont
  {Vel\'azquez-Ahumada}}, \bibinfo {author} {\bibfnamefont {A.}~\bibnamefont
  {Fern\'andez-Prieto}}, \bibinfo {author} {\bibfnamefont {A.}~\bibnamefont
  {Lujambio}}, \bibinfo {author} {\bibfnamefont {V.}~\bibnamefont {Losada}},
  \bibinfo {author} {\bibfnamefont {M.~J.}\ \bibnamefont {Freire}}, \ and\
  \bibinfo {author} {\bibfnamefont {A.}~\bibnamefont {Cabello}},\ }\href
  {\doibase 10.1103/PhysRevLett.116.250404} {\bibfield  {journal} {\bibinfo
  {journal} {Phys. Rev. Lett.}\ }\textbf {\bibinfo {volume} {116}},\ \bibinfo
  {pages} {250404} (\bibinfo {year} {2016})}\BibitemShut {NoStop}%
\bibitem [{\citenamefont {Cerf}\ \emph {et~al.}(1998)\citenamefont {Cerf},
  \citenamefont {Adami},\ and\ \citenamefont {Kwiat}}]{Cerf98}%
  \BibitemOpen
  \bibfield  {author} {\bibinfo {author} {\bibfnamefont {N.~J.}\ \bibnamefont
  {Cerf}}, \bibinfo {author} {\bibfnamefont {C.}~\bibnamefont {Adami}}, \ and\
  \bibinfo {author} {\bibfnamefont {P.~G.}\ \bibnamefont {Kwiat}},\ }\href
  {\doibase 10.1103/PhysRevA.57.R1477} {\bibfield  {journal} {\bibinfo
  {journal} {Phys. Rev. A}\ }\textbf {\bibinfo {volume} {57}},\ \bibinfo
  {pages} {R1477} (\bibinfo {year} {1998})}\BibitemShut {NoStop}%
\bibitem [{\citenamefont {Mermin}(1990)}]{Mermin90}%
  \BibitemOpen
  \bibfield  {author} {\bibinfo {author} {\bibfnamefont {N.~D.}\ \bibnamefont
  {Mermin}},\ }\href {\doibase 10.1103/PhysRevLett.65.1838} {\bibfield
  {journal} {\bibinfo  {journal} {Phys. Rev. Lett.}\ }\textbf {\bibinfo
  {volume} {65}},\ \bibinfo {pages} {1838} (\bibinfo {year}
  {1990})}\BibitemShut {NoStop}%
\bibitem [{\citenamefont {Kochen}\ and\ \citenamefont {Specker}(1967)}]{KS}%
  \BibitemOpen
  \bibfield  {author} {\bibinfo {author} {\bibfnamefont {S.~B.}\ \bibnamefont
  {Kochen}}\ and\ \bibinfo {author} {\bibfnamefont {E.}~\bibnamefont
  {Specker}},\ }\href@noop {} {\bibfield  {journal} {\bibinfo  {journal} {J.
  Math. Mech.}\ }\textbf {\bibinfo {volume} {17}},\ \bibinfo {pages} {59}
  (\bibinfo {year} {1967})}\BibitemShut {NoStop}%
\bibitem [{\citenamefont {Cabello}\ \emph {et~al.}(2010)\citenamefont
  {Cabello}, \citenamefont {Severini},\ and\ \citenamefont {Winter}}]{CSW}%
  \BibitemOpen
  \bibfield  {author} {\bibinfo {author} {\bibfnamefont {A.}~\bibnamefont
  {Cabello}}, \bibinfo {author} {\bibfnamefont {S.}~\bibnamefont {Severini}}, \
  and\ \bibinfo {author} {\bibfnamefont {A.}~\bibnamefont {Winter}},\ }
  %\href{https://arxiv.org/abs/1010.2163} 
{\bibfield
  {journal} {\bibinfo  {journal} {Phys. Rev. Lett.}\ }\textbf {\bibinfo
  {volume} {112}},\ \bibinfo {pages} {040401} (\bibinfo {year}
  {2014})}\BibitemShut {NoStop}%  
%  {arXiv:1010.2163 [quant-ph]}\ } (\bibinfo {year} {2010})}\BibitemShut{NoStop}%
\bibitem [{\citenamefont {Loudon}(1998)}]{Loudon}%
  \BibitemOpen
  \bibfield  {author} {\bibinfo {author} {\bibfnamefont {R.}~\bibnamefont
  {Loudon}},\ }\href@noop {} {\bibfield  {journal} {\bibinfo  {journal} {Phys.
  Rev. A}\ }\textbf {\bibinfo {volume} {58}},\ \bibinfo {pages} {4904}
  (\bibinfo {year} {1998})}\BibitemShut {NoStop}%
\bibitem [{\citenamefont {Scully}\ and\ \citenamefont
  {Zubairy}(1997)}]{QOpticsbook}%
  \BibitemOpen
  \bibfield  {author} {\bibinfo {author} {\bibfnamefont {M.~O.}\ \bibnamefont
  {Scully}}\ and\ \bibinfo {author} {\bibfnamefont {M.~S.}\ \bibnamefont
  {Zubairy}},\ }\href {\doibase 10.1017/CBO9780511813993} {\emph {\bibinfo
  {title} {Quantum Optics}}}\ (\bibinfo  {publisher} {Cambridge University
  Press},\ \bibinfo {year} {1997})\BibitemShut {NoStop}%
\bibitem [{\citenamefont {Fine}(1982)}]{Fine82}%
  \BibitemOpen
  \bibfield  {author} {\bibinfo {author} {\bibfnamefont {A.}~\bibnamefont
  {Fine}},\ }\href {\doibase 10.1103/PhysRevLett.48.291} {\bibfield  {journal}
  {\bibinfo  {journal} {Phys. Rev. Lett.}\ }\textbf {\bibinfo {volume} {48}},\
  \bibinfo {pages} {291} (\bibinfo {year} {1982})}\BibitemShut {NoStop}%
\bibitem [{\citenamefont {Kurzy\ifmmode~\acute{n}\else
  \'{n}\fi{}ski}(2017)}]{Kurzynski2017}%
  \BibitemOpen
  \bibfield  {author} {\bibinfo {author} {\bibfnamefont {P.}~\bibnamefont
  {Kurzy\ifmmode~\acute{n}\else \'{n}\fi{}ski}},\ }\href@noop {} {\bibfield
  {journal} {\bibinfo  {journal} {Phys. Rev. A}\ }\textbf {\bibinfo {volume}
  {95}},\ \bibinfo {pages} {012133} (\bibinfo {year} {2017})}\BibitemShut
  {NoStop}%
\bibitem [{\citenamefont {Cabello}(2013{\natexlab{a}})}]{Exclusivity}%
  \BibitemOpen
  \bibfield  {author} {\bibinfo {author} {\bibfnamefont {A.}~\bibnamefont
  {Cabello}},\ }\href@noop {} {\bibfield  {journal} {\bibinfo  {journal} {Phys.
  Rev. Lett.}\ }\textbf {\bibinfo {volume} {110}},\ \bibinfo {pages} {060402}
  (\bibinfo {year} {2013}{\natexlab{a}})}\BibitemShut {NoStop}%
\bibitem [{\citenamefont {Fritz}\ \emph {et~al.}(2013)\citenamefont {Fritz},
  \citenamefont {Sainz}, \citenamefont {Augusiak}, \citenamefont {Brask},
  \citenamefont {Chaves}, \citenamefont {Leverrier},\ and\ \citenamefont
  {Acin}}]{LO}%
  \BibitemOpen
  \bibfield  {author} {\bibinfo {author} {\bibfnamefont {T.}~\bibnamefont
  {Fritz}}, \bibinfo {author} {\bibfnamefont {A.~B.}\ \bibnamefont {Sainz}},
  \bibinfo {author} {\bibfnamefont {R.}~\bibnamefont {Augusiak}}, \bibinfo
  {author} {\bibfnamefont {J.~B.}\ \bibnamefont {Brask}}, \bibinfo {author}
  {\bibfnamefont {R.}~\bibnamefont {Chaves}}, \bibinfo {author} {\bibfnamefont
  {A.}~\bibnamefont {Leverrier}}, \ and\ \bibinfo {author} {\bibfnamefont
  {A.}~\bibnamefont {Acin}},\ }\href@noop {} {\bibfield  {journal} {\bibinfo
  {journal} {Nat. Commun.}\ }\textbf {\bibinfo {volume} {4}},\ \bibinfo {pages}
  {2263} (\bibinfo {year} {2013})}\BibitemShut {NoStop}%
\bibitem [{\citenamefont {Markiewicz}\ \emph {et~al.}(2014)\citenamefont
  {Markiewicz}, \citenamefont {Kurzy\ifmmode~\acute{n}\else \'{n}\fi{}ski},
  \citenamefont {Thompson}, \citenamefont {Lee}, \citenamefont {Soeda},
  \citenamefont {Paterek},\ and\ \citenamefont {Kaszlikowski}}]{Markiewicz14}%
  \BibitemOpen
  \bibfield  {author} {\bibinfo {author} {\bibfnamefont {M.}~\bibnamefont
  {Markiewicz}}, \bibinfo {author} {\bibfnamefont {P.}~\bibnamefont
  {Kurzy\ifmmode~\acute{n}\else \'{n}\fi{}ski}}, \bibinfo {author}
  {\bibfnamefont {J.}~\bibnamefont {Thompson}}, \bibinfo {author}
  {\bibfnamefont {S.-Y.}\ \bibnamefont {Lee}}, \bibinfo {author} {\bibfnamefont
  {A.}~\bibnamefont {Soeda}}, \bibinfo {author} {\bibfnamefont
  {T.}~\bibnamefont {Paterek}}, \ and\ \bibinfo {author} {\bibfnamefont
  {D.}~\bibnamefont {Kaszlikowski}},\ }\href {\doibase
  10.1103/PhysRevA.89.042109} {\bibfield  {journal} {\bibinfo  {journal} {Phys.
  Rev. A}\ }\textbf {\bibinfo {volume} {89}},\ \bibinfo {pages} {042109}
  (\bibinfo {year} {2014})}\BibitemShut {NoStop}%
\bibitem [{\citenamefont {Cabello}\ \emph {et~al.}(2013)\citenamefont
  {Cabello}, \citenamefont {Danielsen}, \citenamefont {L\'opez-Tarrida},\ and\
  \citenamefont {Portillo}}]{CabelloBasicGraphs}%
  \BibitemOpen
  \bibfield  {author} {\bibinfo {author} {\bibfnamefont {A.}~\bibnamefont
  {Cabello}}, \bibinfo {author} {\bibfnamefont {L.~E.}\ \bibnamefont
  {Danielsen}}, \bibinfo {author} {\bibfnamefont {A.~J.}\ \bibnamefont
  {L\'opez-Tarrida}}, \ and\ \bibinfo {author} {\bibfnamefont {J.~R.}\
  \bibnamefont {Portillo}},\ }\href {\doibase 10.1103/PhysRevA.88.032104}
  {\bibfield  {journal} {\bibinfo  {journal} {Phys. Rev. A}\ }\textbf {\bibinfo
  {volume} {88}},\ \bibinfo {pages} {032104} (\bibinfo {year}
  {2013})}\BibitemShut {NoStop}%
\bibitem [{\citenamefont {Cabello}(2012)}]{CabelloMermin}%
  \BibitemOpen
  \bibfield  {author} {\bibinfo {author} {\bibfnamefont {A.}~\bibnamefont
  {Cabello}},\ }\href {https://arxiv.org/abs/1212.1756} {\bibfield  {journal}
  {\bibinfo  {journal} {arXiv:1212.1756 [quant-ph]}\ } (\bibinfo {year}
  {2012})}\BibitemShut {NoStop}%
\bibitem [{\citenamefont {Short}\ and\ \citenamefont
  {Barrett}(2010)}]{Short10}%
  \BibitemOpen
  \bibfield  {author} {\bibinfo {author} {\bibfnamefont {A.~J.}\ \bibnamefont
  {Short}}\ and\ \bibinfo {author} {\bibfnamefont {J.}~\bibnamefont
  {Barrett}},\ }\href {\doibase 10.1088/1367-2630/12/3/033034} {\bibfield
  {journal} {\bibinfo  {journal} {New J. Phys.}\ }\textbf {\bibinfo {volume}
  {12}},\ \bibinfo {pages} {033034} (\bibinfo {year} {2010})}\BibitemShut
  {NoStop}%
\bibitem [{\citenamefont {Streater}(2000)}]{Streater00}%
  \BibitemOpen
  \bibfield  {author} {\bibinfo {author} {\bibfnamefont {R.~F.}\ \bibnamefont
  {Streater}},\ }\href {\doibase 10.1063/1.533322} {\bibfield  {journal}
  {\bibinfo  {journal} {J. Math. Phys.}\ }\textbf {\bibinfo {volume} {41}},\
  \bibinfo {pages} {3556} (\bibinfo {year} {2000})}\BibitemShut {NoStop}%
\bibitem [{\citenamefont {Cabello}(2013{\natexlab{b}})}]{CabelloSimple}%
  \BibitemOpen
  \bibfield  {author} {\bibinfo {author} {\bibfnamefont {A.}~\bibnamefont
  {Cabello}},\ }\href {\doibase 10.1103/PhysRevLett.110.060402} {\bibfield
  {journal} {\bibinfo  {journal} {Phys. Rev. Lett.}\ }\textbf {\bibinfo
  {volume} {110}},\ \bibinfo {pages} {060402} (\bibinfo {year}
  {2013}{\natexlab{b}})}\BibitemShut {NoStop}%
\bibitem [{\citenamefont {Hammack}\ \emph {et~al.}(2011)\citenamefont
  {Hammack}, \citenamefont {Imrich},\ and\ \citenamefont {Klavzar}}]{ORP}%
  \BibitemOpen
  \bibfield  {author} {\bibinfo {author} {\bibfnamefont {R.}~\bibnamefont
  {Hammack}}, \bibinfo {author} {\bibfnamefont {W.}~\bibnamefont {Imrich}}, \
  and\ \bibinfo {author} {\bibfnamefont {S.}~\bibnamefont {Klavzar}},\
  }\href@noop {} {\emph {\bibinfo {title} {Handbook of Product Graphs}}}\
  (\bibinfo  {publisher} {CRC Press, Boca Raton, FL},\ \bibinfo {year}
  {2011})\BibitemShut {NoStop}%
\bibitem [{\citenamefont {Klyachko}\ \emph {et~al.}(2008)\citenamefont
  {Klyachko}, \citenamefont {Can}, \citenamefont
  {Binicio\ifmmode~\breve{g}\else \u{g}\fi{}lu},\ and\ \citenamefont
  {Shumovsky}}]{KCBS}%
  \BibitemOpen
  \bibfield  {author} {\bibinfo {author} {\bibfnamefont {A.~A.}\ \bibnamefont
  {Klyachko}}, \bibinfo {author} {\bibfnamefont {M.~A.}\ \bibnamefont {Can}},
  \bibinfo {author} {\bibfnamefont {S.}~\bibnamefont
  {Binicio\ifmmode~\breve{g}\else \u{g}\fi{}lu}}, \ and\ \bibinfo {author}
  {\bibfnamefont {A.~S.}\ \bibnamefont {Shumovsky}},\ }\href {\doibase
  10.1103/PhysRevLett.101.020403} {\bibfield  {journal} {\bibinfo  {journal}
  {Phys. Rev. Lett.}\ }\textbf {\bibinfo {volume} {101}},\ \bibinfo {pages}
  {020403} (\bibinfo {year} {2008})}\BibitemShut {NoStop}%
\end{thebibliography}

\end{document}